\documentclass{article}

\usepackage[a4paper, left=3cm, right=3cm]{geometry}
\usepackage{amsmath}
\usepackage{amssymb}
\usepackage{amsthm}
\usepackage{graphicx}
\usepackage{caption}
\usepackage{subcaption}
\usepackage{authblk}
\usepackage{bbm}
\usepackage{comment}
\usepackage[numbers,sort]{natbib}
\usepackage[colorlinks=true, allcolors=blue]{hyperref}
\usepackage{xcolor}
\usepackage{ulem}
\usepackage[usestackEOL]{stackengine}
\usepackage{doi}

\theoremstyle{definition}

\newtheorem{remark}{Remark}

\title{Approximating particle-based clustering dynamics by stochastic PDEs}

\author[1,2]{Nathalie Wehlitz}
\author[2]{Mohsen Sadeghi}
\author[1]{Alberto Montefusco}
\author[1,2]{Christof Schütte}
\author[3]{Grigorios A. Pavliotis}
\author[1]{Stefanie Winkelmann\footnote{Email of corresponding author: \href{mailto:winkelmann@zib.de}{winkelmann@zib.de}}}

\affil[1]{Zuse Institute Berlin, Berlin, Germany}
\affil[2]{Free University Berlin, Berlin, Germany}
\affil[3]{Imperial College London, London, UK}

\date{}

\begin{document}

\maketitle

\begin{abstract}
This work proposes stochastic partial differential equations (SPDEs) as a practical tool to replicate clustering effects of more detailed particle-based dynamics. Inspired by membrane-mediated receptor dynamics on cell surfaces, we formulate a stochastic particle-based model for diffusion and pairwise interaction of particles, leading to intriguing clustering phenomena.
Employing numerical simulation and cluster detection methods, we explore the approximation of the particle-based clustering dynamics through mean-field approaches. We find that SPDEs successfully reproduce spatiotemporal clustering dynamics, not only in the initial cluster formation period, but also on longer time scales where the successive merging of clusters cannot be tracked by deterministic mean-field models.  The computational efficiency of the SPDE approach allows us to generate extensive statistical data for parameter estimation in a simpler model that uses a Markov jump process to capture the temporal evolution of the cluster number.

\quad 

\end{abstract}

\section{Introduction}

Biomolecules like lipids, proteins, and nucleic acids are crucial for cell functions, organizing spatially through complex interactions. Depending on selectivity and strength relative to thermal noise, these interactions can lead to aggregation, cluster formation, and phase separation~\cite{kockelkoren2023}. 
Cell membranes provide a convenient substrate for such phenomena, with membrane-bound proteins utilizing a variety of direct and indirect interactions to form clusters on the cell surface~\cite{weikl2018, fournier2024,johannes2018}. In a prominent example, in vitro studies of living neural cells have shown that receptors, such as $\mu$\textit{-opioid receptors}, form clusters on the cell surface~\cite{moeller2020}. 
Although high-resolution molecular models might be needed to fully understand the detailed mechanisms of this and similar phenomena, large-scale simulations with particles that each represent one or several molecules can provide physics-based insights into how membrane-mediated interactions lead to cluster formation. In a recent approach~\cite{sadeghi2021, sadeghi2022}, bilayer membrane patches were modeled, interacting with a collection of peripheral proteins, which laterally diffuse and induce local curvature in the membrane. This particle-based membrane model successfully reproduces cluster formation under the influence of purely membrane-mediated interactions.
Beyond biochemical applications, clustering also appears in biological or social systems where particles refer to individuals that exhibit collective behavior. This includes phenomena such as swarming of animals or the formation of consensus in opinion dynamics~\cite{garnier2016, mogilner2003, topaz2008, leverentz2009, bicego2024computationcontrolunstablesteady}. 

Generally, individual-based models are complex to solve and computationally intensive. 
Scaling them up while preserving spatiotemporal information and maintaining stochastic effects, which are crucial for cluster formation, is therefore a central challenge. In this article, we investigate the extent to which mean-field approximations can reproduce the clustering behavior observed in stochastic particle-based models. Our central goal is to demonstrate that stochastic partial differential equations are a practical tool for efficiently simulating and studying clustering phenomena in biological systems and other applications. 

We begin by formulating a stochastic particle-based model that may be viewed as a reduced and abstracted version of the biochemical membrane model mentioned above~\cite{sadeghi2021, sadeghi2022}. 
The interaction potential, which will be chosen as the Morse potential~\cite{carrillo2019, dorsogna2006}, is parameterized to include attractive forces that cause particles to coalesce and short-range repulsion that accounts for space exclusion, thereby weakening attraction. The particle-based model forms a high-dimensional system of stochastic differential equations (SDEs), with the number of equations equal to the number of configurational degrees of freedom of the particle system. 
This makes numerical integration computationally intensive for large population sizes, motivating the formulation of mean-field approximations in form of a partial differential equation~(PDE), 
initially studied by McKean~\cite{mckean1966}, or in form of a stochastic partial differential equation~(SPDE), 
introduced by Dean and Kawasaki ~\cite{dean1996,kawasaki1998}. 
The resulting \textit{Dean--Kawasaki equation} describes fluctuations in the density of diffusing particles in the regime of large (but finite) particle numbers.
This SPDE is a highly singular equation whose solution theory is a subject of ongoing research~\cite{konarovskyi2019,konarovskyi2020,djurdjevac2022weak, fehrmanngess2024}. We therefore emphasize that here we consider the SPDE as a \textit{modeling tool} by using it in its regularized form~\cite{cornalba2023, cornalba_fischer_ingmanns2023}. 
In~\cite{konarovskyi2024reversible} and \cite{andres2010particle}, Dean-Kawasaki-type SPDEs are used to study reversible fragmentating-coagulating processes of particles with local interactions.
The underlying concept of fluctuating hydrodynamics has also found applications beyond physics-based models, e.g.,~in the context of interacting agent systems~\cite{djurdjevac2022feedback,helfmann2021interacting,kim2017stochastic}. 

Given the modeling approaches, we aim to analyze and compare the clustering behavior resulting from the induced spatiotemporal dynamics. Generally, one can distinguish two stages of clustering dynamics:
\begin{enumerate}
    \item The \textit{initial period of cluster formation} after starting, e.g., in uniform distribution. Clusters are created and change their width and shape based on an interplay of diffusion and interaction of the particles. 
    \item The \textit{long-term clustering dynamics} after initial cluster formation. The positions of the cluster centers change over time due to spatial movement of particles. 
    Through mutual interactions and random spatial encounters, clusters gradually merge. Occasionally, with a small probability, they split again. This dynamic process leads to changes in the number and composition of clusters. 
\end{enumerate}
As for the mean-field PDE, the initial period of cluster formation has been studied analytically based on linear stability analysis~\cite{garnier2016,delfau2016, Chazelle_al2017a}, providing estimates for the number of clusters and the time to cluster formation, and showing phase transitions in dependence on the strength of interaction~\cite{carrillo2020}. Once clusters have formed, their total number and individual positions are an invariant of the deterministic system of motion. To continue with the second stage of clustering dynamics, stochastic effects are required. In this work, we focus on the long-term stochastic clustering dynamics after the initial period of cluster formation, demonstrating that the clustering behavior of the SPDE aligns closely with that of the particle-based model. Individual numerical simulations of the two models are used to show agreement in the shape and positional movement of the clusters. Also first- and second-order moments of cluster counts and their statistical distribution are compared, finding concordance across all time scales.
These insights are then used to (i) explore the long-term statistics of cluster counts by means of long-term simulations of the SPDE, and (ii) to perform further coarse-graining by averaging out all spatial resolution and using a Markov jump process to describe the temporal evolution of the number of clusters. 
Although this Markovian cluster-counting process ignores the intrinsic memory effects of the spatially resolved models, it shows a surprisingly strong agreement with the SPDE model regarding the distribution of cluster counts, and consequently with particle-based dynamics as well. 
The numerical efficiency of the SPDE enables the estimation of the jump rates for the cluster-counting process, which would be prohibitively expensive using particle-based dynamics.

\quad

The paper is structured as follows. In Section~\ref{sec:Modeling approaches} we introduce the stochastic particle-based interaction model and its approximation by mean-field approaches. Exemplary simulation runs are shown to get a first impression of the clustering behavior, and some well-known findings about the cluster formation period are summarized. In Section~\ref{sec:Clustering dynamics} we compare the long-term stochastic clustering dynamics of the particle-based model and the SPDE at qualitative and quantitative levels. Finally, we use the SPDE to estimate the jump rates of the Markovian cluster-counting process. 

\section{Model setup} 
\label{sec:Modeling approaches}

In the scenario considered, the dynamics is mass-conserving and non-reactive, unlike biochemical reaction-diffusion systems where particle numbers change due to reactions~\cite{winkelmann2020stochastic}. For membrane-mediated receptor kinetics, assuming non-reactivity is reasonable because receptor numbers remain constant within the relevant time frames. This constancy simplifies formulating a stochastic particle-based model compared to scenarios with varying molecule numbers, where probabilistic modeling is more complex~\cite{del2023chemical,del2022probabilistic}. We use periodic boundary conditions, implying that the domain of the system is much larger than the region we consider, and that curvature or edge effects are negligible. These assumptions are justified by the size difference between living cells (tens of micrometers) and the individual receptors modeled (a few nanometers).

We introduce the particle-based model in Section~\ref{subsec:Stochastic particle-based model} using a path-wise formulation, and we define the mean-field approximations in Section~\ref{sec:mean-field}. The formulations are limited to one-dimensional domains of motion, but extensions to higher dimensions are straightforward. In Section~\ref{sec:numerical_experiment}, an exemplary numerical simulation gives a first impression of the stochastic clustering dynamics, while in Section~\ref{PDE} we provide a short summary of PDE-analytics for the initial cluster formation period. 

\subsection{Stochastic particle-based model}
\label{subsec:Stochastic particle-based model}

We consider $N\in \mathbb{N}$ particles moving on the one-dimensional torus $\mathbb{T} = \big[-\frac{L}{2},\frac{L}{2}\big]$, $L>0$. Let $\boldsymbol{X}(t)=(X_1(t),...,X_N(t))$ denote the (random) system state at time $t\in [0,\infty)$, where $X_i(t)\in \mathbb{T}$ is the position of particle $i\in\{1,...,N\}$. The positions evolve according to the set of coupled SDEs
\begin{equation}\label{eq:PBD}
    dX_i(t) =  
     - \frac{1}{N}\sum_{j=1}^N F'\big(X_i(t)-X_j(t)\big) \, dt + \sigma dW_i(t), \quad i=1,...,N,
\end{equation}
for $F'(x)=\frac{d}{dx}F(x)$, where $F:\mathbb{R}\to \mathbb{R}$ is a potential modeling pairwise interactions between particles, 
$W_1, \dots, W_N$ are independent standard Wiener processes in $\mathbb{R}$. Here, $\sigma > 0$ is the noise strength, which is assumed constant, as is standard when modeling proteins in membranes.
The stochastic dynamics given by~\eqref{eq:PBD} is henceforth referred to as the \textit{particle-based dynamics}.

Let the interaction potential be given by the generalized \textit{Morse potential}~\cite{carrillo2019, dorsogna2006} 
\begin{equation}\label{eq:F}
    F(x)=-C_a \exp\left(-\frac{|x|}{l_a}\right) + C_r \exp\left(-\frac{|x|}{l_r}\right),
\end{equation}
where $l_a, l_r > 0$ represent the length scales of the attractive and repulsive forces, respectively, and $C_a, C_r > 0$ are the corresponding strengths. Since the motion space for the particles is assumed to be a torus, it is reasonable to use a periodic form of~\eqref{eq:F}, which we will do in the following. For a definition and derivation, we refer to~\cite{bandegi2018}.

We found the Morse potential given in \eqref{eq:F} provides a good approximation of pairwise membrane-mediated interactions measured in~\cite{sadeghi2022}. In general, the Morse potential encapsulates the soft-core space exclusion between membrane proteins at close range, as well as the long-range attraction resulting from, e.g., fluctuation-mediated interactions~\cite{weikl2019}.
Apart from this, the Morse potential has also proven useful for modeling attraction and repulsion in swarming dynamics or opinion formation~\cite{mogilner2003, topaz2008, leverentz2009}.

\subsubsection*{Clustering framework} We will consider the case of $\frac{l_r}{l_a}<1$ and $\frac{C_r}{C_a}<1$, which implies that attraction always dominates repulsion~\cite{mogilner2003}, although at close distances the attractive force diminishes due to the presence of the repulsive force. In addition, the attraction and repulsion ranges, $l_a$ and $l_r$, must be significantly smaller than the spatial domain length $L$ to ensure that interactions remain sufficiently localized. 
Moreover, we assume that the noise strength $\sigma$ is less than the critical noise strength, to ensure that the uniform state becomes unstable and that phase transitions occur, see~\cite[Thm. 5.11]{carrillo2020} and ~\cite{bicego2024computationcontrolunstablesteady, bertoli2024}. At high noise strengths, diffusion dominates and the particle dynamics behaves like a system of $N$ independent Brownian motions, converging to the uniform steady state~\cite{garnier2016}.

\subsection{Mean-field approximations}\label{sec:mean-field}

For large population sizes $N$, the particle-based dynamics may be approximated by a mean-field PDE or SPDE. These equations describe the spatiotemporal evolution of the concentration $c(x,t)$ of particles, depending on position $x\in \mathbb{T}$ and time $t \geq 0$, and approximate the empirical measure of the particle-based dynamics in the case of large $N$~\cite{dawson1983, gaertner1988}.  

For the system under consideration, the \textit{mean-field partial differential equation (PDE)} is given by 
\begin{align}\label{eq:PDE mean-field_1d}
    \partial_t c(x,t) = \partial_x (c(x,t) (F' * c(\cdot, t))(x)) + \frac{\sigma^2}{2}  \partial_{xx} c(x,t),
\end{align}
where 
\begin{equation}
    (F'* c(\cdot, t))(x):= \int_{\mathbb{T}} F'(x-y)c(y,t)\, dy
\end{equation}
is the convolution between the interaction force $F'$ and the density $c$. The partial integro-differential equation~\eqref{eq:PDE mean-field_1d} is a nonlinear, non-local Fokker-Planck equation that is also called \textit{McKean-Vlasov PDE} or \textit{aggregation-diffusion equation}.\footnote{We note that the \textit{McKean-Vlasov PDE}~\eqref{eq:PDE mean-field_1d} differs from the \textit{McKean SDE}, which  describes the motion of a single particle in the limit $N \to \infty$ where the particles become independent by propagation of chaos~\cite{mckean1966}.  The distribution of a solution to the McKean SDE is also a solution to our mean-field PDE.}
 
The PDE~\eqref{eq:PDE mean-field_1d} approximates the particle-based dynamics for $N\to \infty$, but it lacks stochastic effects crucial for systems of finite size $N$, especially in the case of clustering behavior. This randomness is maintained by the corresponding \textit{stochastic partial differential equation (SPDE)}
\begin{align} \label{eq:Dean-Kawasaki}
    \partial_t c(x,t) = \partial_x (c(x,t) (F' * c(\cdot, t))(x)) + \frac{\sigma^2}{2}  \partial_{xx} c(x,t) + \frac{\sigma}{\sqrt{N}}\partial_x (\sqrt{c(x,t)}\xi(x,t)),
\end{align}
also called \textit{Dean--Kawasaki equation}~\cite{dean1996,kawasaki1998}. Here, $\xi(x,t)$ denotes space-time white noise, i.e.,
\begin{equation}
 \mathbb{E}\left(\xi(x,t)\right)=0, \quad  \quad  \mathbb{E}\left(\xi(x,t)\xi(x',t')\right)= \delta(x-x')\delta(t-t'),   \quad \quad \forall t,t'\geq 0, \forall x,x'\in  \mathbb{T},
\end{equation}
where $\delta(x)$ denotes the Dirac delta distribution. 

Originally, it was demonstrated that the empirical density corresponding to the particle-based dynamics~\eqref{eq:PBD} solves the Dean--Kawasaki equation, indicating that the SPDE does not contain different information than the particle-based model itself~\cite{dean1996}. In fact, it was shown in~\cite{konarovskyi2019} that there is no other martingale solution than the empirical density. How the highly singular equation can be interpreted apart from the empirical density is a subject of ongoing research; we refer to~\cite{djurdjevac2022weak} for a detailed discussion. 

We focus here on a version of the \textit{regularized} solution, that is, a spatially discretized and therefore implicitly noise truncated solution to the SPDE~\eqref{eq:Dean-Kawasaki}. In~\cite{cornalba2023, cornalba_fischer_ingmanns2023}, it was demonstrated that discretizing the SPDE yields a meaningful approximation of the particle-based dynamics, both in the non-interacting case and in the case of pairwise interaction. The approximation quality is limited by the spatial discretization error and the error associated with the negative part of the discretized solution. Sensitivity regarding the choice of the spatial discretization step is therefore required. Particularly in the case of clustering behavior, there will be regions of low concentration where a discretized solution of the SPDE can become negative. The extent to which the error associated with the negative part can be controlled is analyzed in~\cite{cornalba_fischer_ingmanns2023}.

For a large number $N$ of particles and in case of a low-dimensional space $\mathbb{T}$ of motion, a numerical discretization of the SPDE~\eqref{eq:Dean-Kawasaki} can be computationally much cheaper than the particle-based simulation. This is the case if the spatial discretisation parameter $h$ (referring to the grid size) is significantly larger than $1/N$, that is, if there is on average more than one particle per grid cell~\cite{cornalba_fischer_ingmanns2023}. 
The efficiency of the SPDE-simulation will be useful for long-term simulations of the clustering dynamics that will be done in Section~\ref{sec:Clustering dynamics}. 

\subsection{Exemplary numerical simulation} \label{sec:numerical_experiment}

\begin{figure}
    \centering
    \begin{subfigure}{0.32\textwidth}
        \includegraphics[width=\textwidth]{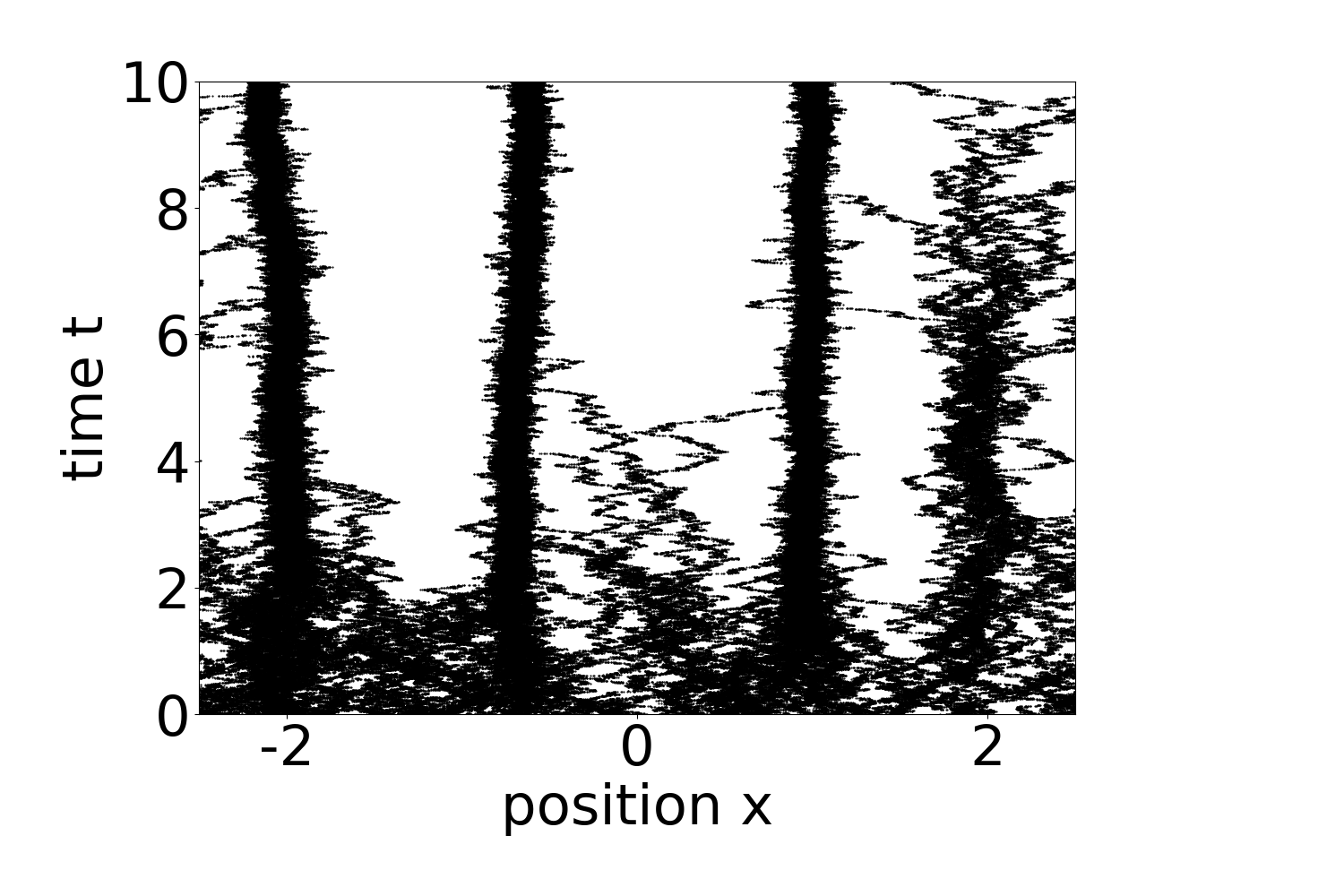}
        \caption{\centering particle-based trajectories \newline \,}
        \label{fig:PBD_compared_SPDE_PBDtraject}
    \end{subfigure}
       \begin{subfigure}{0.32\textwidth}
        \includegraphics[width=\textwidth]{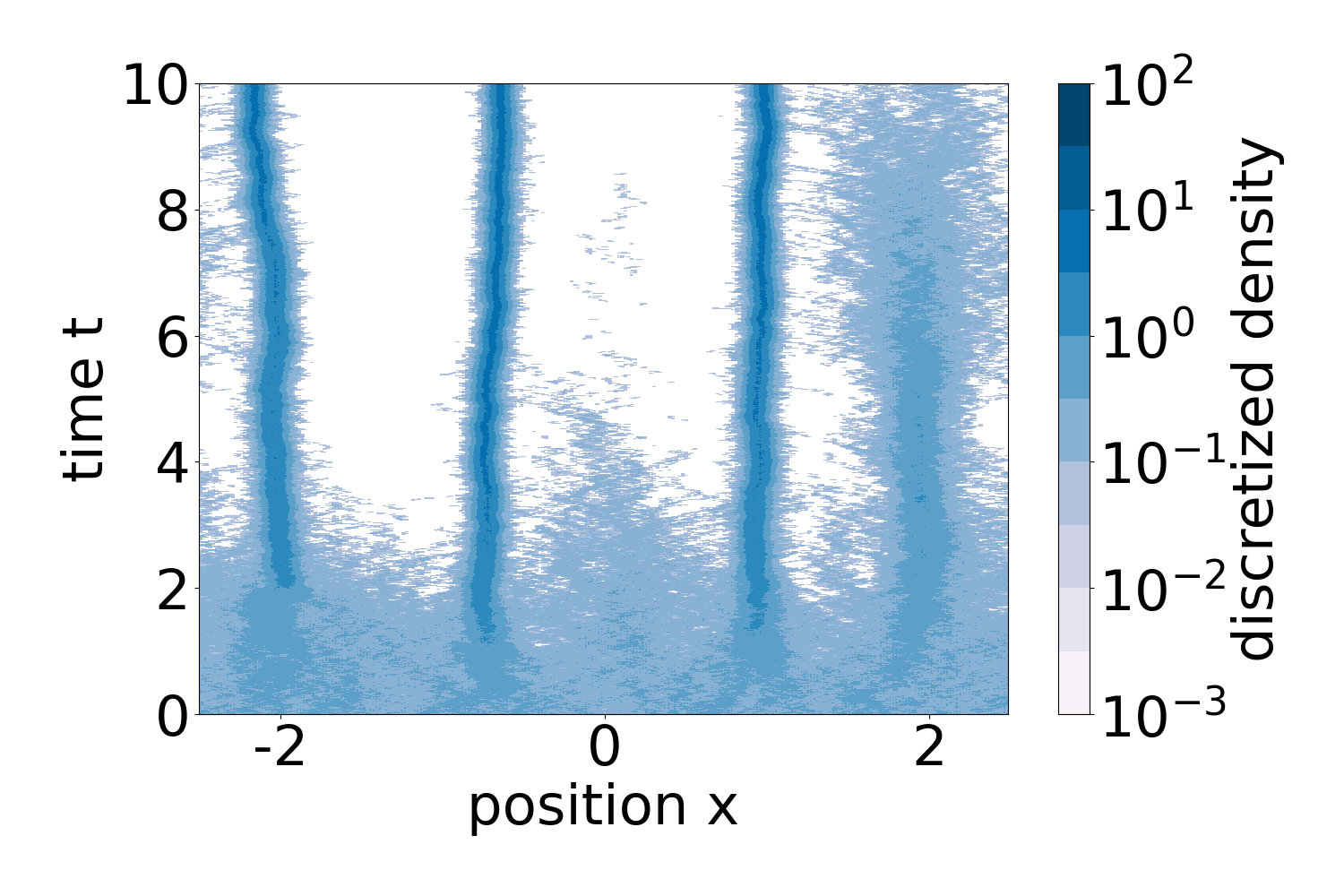}
        \caption{\centering empirical distribution \newline of particle positions}
        \label{fig:PBD_compared_SPDE_PBDhist}
    \end{subfigure}
        \begin{subfigure}{0.32\textwidth}
        \includegraphics[width=\textwidth]{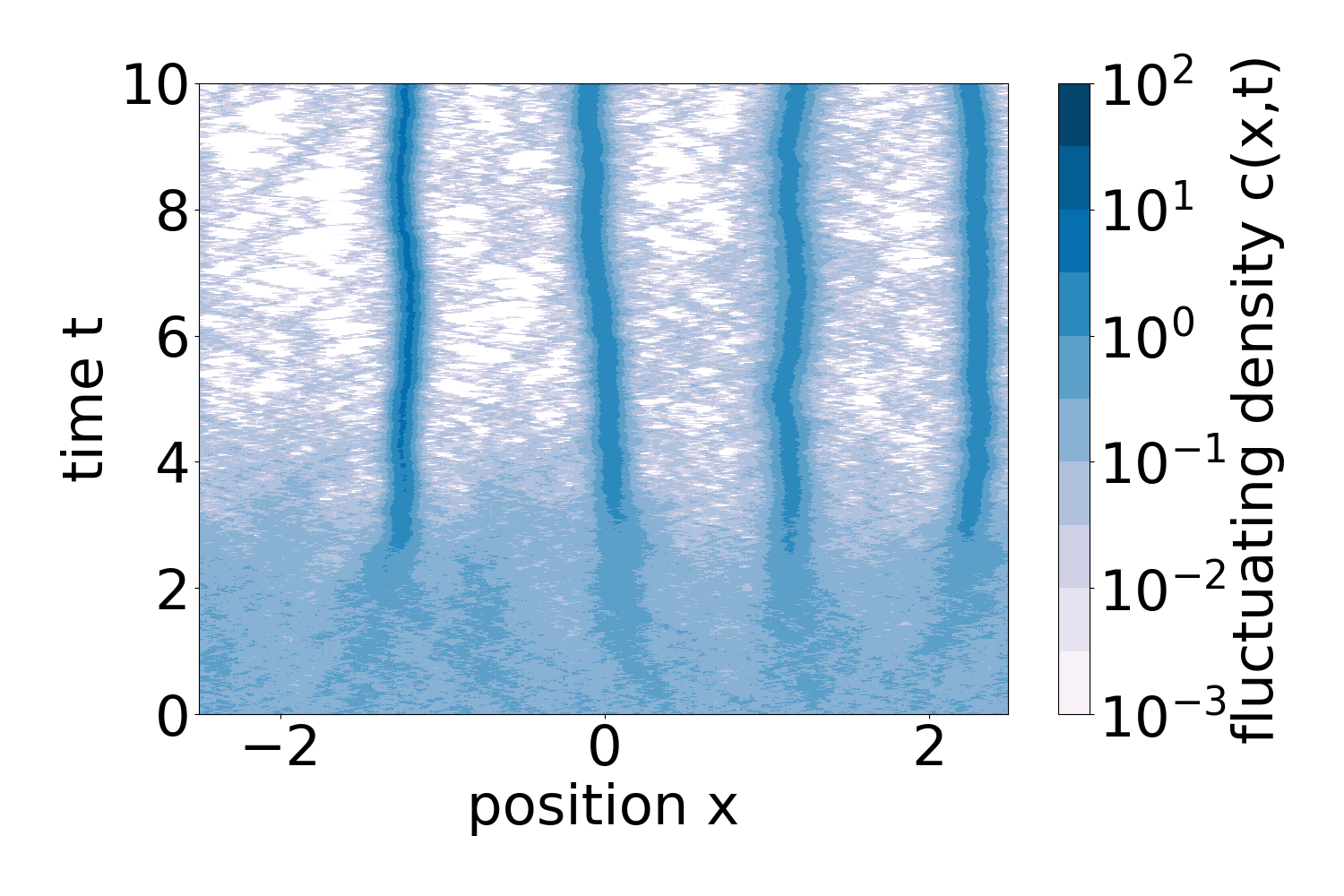}
        \caption{\centering numerical SPDE solution \newline \, }
        \label{fig:PBD_compared_SPDE_SPDE}
    \end{subfigure}
    \caption{\textbf{Individual numerical simulations of initial cluster formation period.} Stochastic dynamics of $N=10^3$ particles on the torus of length $L=5$ until time $t=10$, after starting at time $t=0$ with a uniform distribution.  (a)~spatiotemporal trajectories of the particle-based dynamics~\eqref{eq:PBD}; (b)~empirical distribution for particle-based dynamics; (c)~simulation of the Dean--Kawasaki equation \eqref{eq:Dean-Kawasaki}. For clarity, (a) shows only $10^2$ of the $10^3$ trajectories, while (b) is the empirical distribution of all trajectories. 
    Due to the logarithmic scale, values smaller than $10^{-3}$ (especially negative values for the SPDE) are displayed in white in (b) and (c). Please note: The two simulations, particle-based and Dean-Kawasaki-based, are subject to different realizations of the respective noise process  and thus cannot be compared directly.} 
    \label{fig:PBD_compared_SPDE}
\end{figure}

For our numerical experiments and the results plotted in Figures~\ref{fig:PBD_compared_SPDE}-\ref{fig:number_cluster_average_methods} we choose the following parameter values for the potential function $F$ defined in~\eqref{eq:F}, 
\begin{equation} \label{eq:parameter_val}
  L=5, \quad  C_a=4, \quad  l_a=\frac{1}{40}L, \quad C_r=1, \quad l_r=\frac{1}{100}L,
\end{equation}
as well as a noise strength of $\sigma=0.4$. This noise strength is below the critical noise strength~\cite[Sec. III]{martzel2001} and we are in the clustering framework specified in Section~\ref{subsec:Stochastic particle-based model}, which enables the formation of several clusters. We emphasize the fact that, as long as we are below the critical noise strength, clusters will form. The clustering time and their number depends, however, on the choice of the parameter values. 

Simulations of the particle-based model~\eqref{eq:PBD} are performed using the Euler-Maruyama scheme. We use periodic boundary conditions by manually placing positions in the interval $\mathbb{T}=\big[-\frac{L}{2},\frac{L}{2}\big]$ according to the transformation $
x \mapsto \left(x+\frac{L}{2}\right) \text{mod}\, L - \frac{L}{2}$.  

Regarding the SPDE, we adopt the finite difference scheme utilized in~\cite{cornalba2023} for simulating a Dean--Kawasaki equation without particle interaction, and adapt it to our scenario. 
In the cited work, the authors discretize time by a (two-step) \textit{BDF2 scheme} combined, in the first time step, with a mixed implicit-explicit Euler scheme for the deterministic diffusion and an explicit treatment for the noise. In case of interaction, a non-linear equation has to be solved in each time step, which makes this method computationally intensive. This is why, for the deterministic part, we replaced the scheme by the explicit midpoint method, while for the noise term we adopt the explicit treatment from the first time step of~\cite{cornalba2023}, see Appendix~\ref{app:Num_SPDE} for details. 
With regard to the means and standard deviations over several SPDE-simulations, this method provides similarly good results as the BDF2 scheme. 

For all simulations a time step of $dt=0.001$ is used for both stochastic models, as well as a spatial grid size of $h=L \cdot 2^{-8}$ for the SPDE-simulations; and a bin size of $\tilde{dx}=L \cdot 2^{-8}$ for the discretized density of the particle-based dynamics. 

Figure~\ref{fig:PBD_compared_SPDE} shows a numerical simulation of the particle-based model~\eqref{eq:PBD} compared to a statistically independent simulation of the Dean--Kawasaki equation~\eqref{eq:Dean-Kawasaki}. We observe a qualitative agreement of the two stochastic models in the sense that, after a short initial period, four clusters have formed from the initial uniform starting distribution. In contrast, the PDE would not leave the uniform distribution at all, as this is a stationary solution for the given parameters. Further numerical experiments demonstrate that the conformity between the stochastic models persists when the initial distribution or parameter values are altered. However, due to stochastic effects and independence of the simulations, the positions of the clusters' centers naturally differ for the two models' simulations. Also, the number of clusters formed after the initial period and the time taken for clusters to merge in the course of time vary randomly across different realizations for both models. This means that a comparison of the two stochastic models can be done only on the basis of statistics and not on the basis of individual stochastic realizations. Such a statistical comparison is conducted in Section~\ref{sec:Clustering dynamics}. Before that, we shortly summarize insights about clustering dynamics induced by deterministic dynamics.  

\subsection{Cluster formation for classical mean-field approximation}\label{PDE}

Although the PDE-solution~\eqref{eq:PDE mean-field_1d}, unlike the stochastic dynamics, remains in the uniform distribution once it starts there, changes to the initial distribution can also lead to cluster formation for the deterministic dynamics. 
Particularly for small perturbations of the uniform starting distribution, the number of clusters formed by the PDE-dynamics may initially match that of the stochastic models.
This fact is used in~\cite{garnier2016} for the Hegselmann-Krause model and in~\cite{martzel2001} for the Gaussian attractive interaction using linear stability analysis in order to analyze the formation of clusters for the stochastic dynamics.

This work contains estimations of the number of clusters and the time to cluster formation, as well as conditions for consensus convergence where the dynamics end up in one cluster. Also in~\cite{delfau2016}, linear stability analysis is performed for the case of a generalized version of the Morse potential (containing a general exponent $\alpha$ in the exponential function) in order to derive conditions for initial cluster formation and to make statements about the number and shape of clusters.  

\begin{remark}[Application of linear stability analysis] 
\label{rem:LSA} 
    Applied to our setting, 
    the results from~\cite{garnier2016,Greg2024} deliver an initial cluster formation time of $t_{\text{clu}}\approx 3.05$ and an initial number of clusters of ${n_{\text{clu}}\approx 4.18}$. These results agree well with our observations in Figure~\ref{fig:PBD_compared_SPDE}. To apply the linear stability analysis from~\cite{garnier2016}, we use an interaction force with compact support by cutting off the interaction potential $F'$ at $x_{\text{cut}} = 1.1$, where $F'(x) < 10^{-2}$ for $|x| > x_{\text{cut}}$. This cut-off is used only for the numerical implementation of the stability analysis (and not for any other simulation in this paper) and does not affect the overall dynamics, as $F'$ beyond $x_{\text{cut}}$ is negligible.
\end{remark}

\subsubsection*{Stationary solutions and free energy}
The mean-field PDE~\eqref{eq:PDE mean-field_1d}  is a $2$-Wasserstein gradient flow with respect to the free energy functional which is defined as the sum of the entropy and interaction energy~\cite{carrillo2020},
\begin{equation*}
    \mathcal{F}(c)= \frac{\sigma^2}{2}\int_{\mathbb{T}}c(x) \log (c(x))\, dx + \frac{1}{2} \int_{\mathbb{T}}(F*c)(x)c(x) \, dx
\end{equation*}
for the density $c$. Stationary solutions of the PDE correspond to critical points of the free energy, and they can be calculated by solving the Kirkwood-Monroe integral equation~\cite{bicego2024computationcontrolunstablesteady}. While the uniform distribution is always a stationary solution of the McKean-Vlasov PDE on the torus and in the absence of a confining potential, additional stationary states might appear at sufficiently low temperatures (small $\sigma$), or equivalently for sufficiently strong interactions. In particular, for the interaction potentials that we consider in this paper, at high interaction strengths the uniform state becomes unstable and a second (translation invariant), stable one-cluster stationary state emerges. For our choice of the interaction potential and for the parameter values that we consider--e.g. an interaction potential that is not $H-$stable and whose interaction part is sufficiently localized--this corresponds to a discontinuous (first-order) phase transition~\cite[Prop. 6.2]{carrillo2020}, see also~\cite{Greg2024}. 
Our stochastic simulation results suggest that there are several metastable regions in the free energy landscape in our setting. With the initial period of cluster formation, both the dynamics of the particle-based model and the SPDE reach such a metastable state, corresponding to a multi-clustered state. In the second stage of clustering dynamics (cluster movement and successive merging), a cascade in the free-energy landscape takes place. In the following Section~\ref{sec:Clustering dynamics}, we investigate these metastable stochastic clustering dynamics by means of numerical simulations and statistics.

\section{Stochastic clustering dynamics} \label{sec:Clustering dynamics}

The simulated trajectories depicted in Figure~\ref{fig:PBD_compared_SPDE} show the clustering dynamics over a relatively short time interval. To understand and compare the clustering behavior after initial cluster formation period, we will now investigate longer time intervals. After a qualitative comparison of exemplary trajectories in Section~\ref{sec:qual}, we will compare the statistics of cluster counts over time.
To this end, we define a cluster indirectly using the numerical cluster detection method introduced in Section~\ref{subsec:Cluster detection method}. 
By means of this method, we compare the time-dependent distribution of the cluster numbers, including mean and variances, for the two stochastic models in Section~\ref{sec:statistics}. Finally, in Section~\ref{sec:MJP}, we further approximate the temporal evolution of cluster counts by a Markov jump process, using long-term simulations of the SPDE for estimating the jump rates.

\subsection{Qualitative spatial comparison} \label{sec:qual}

\begin{figure}
    \centering
    \begin{subfigure}{0.49\textwidth}
    \includegraphics[width=\textwidth]{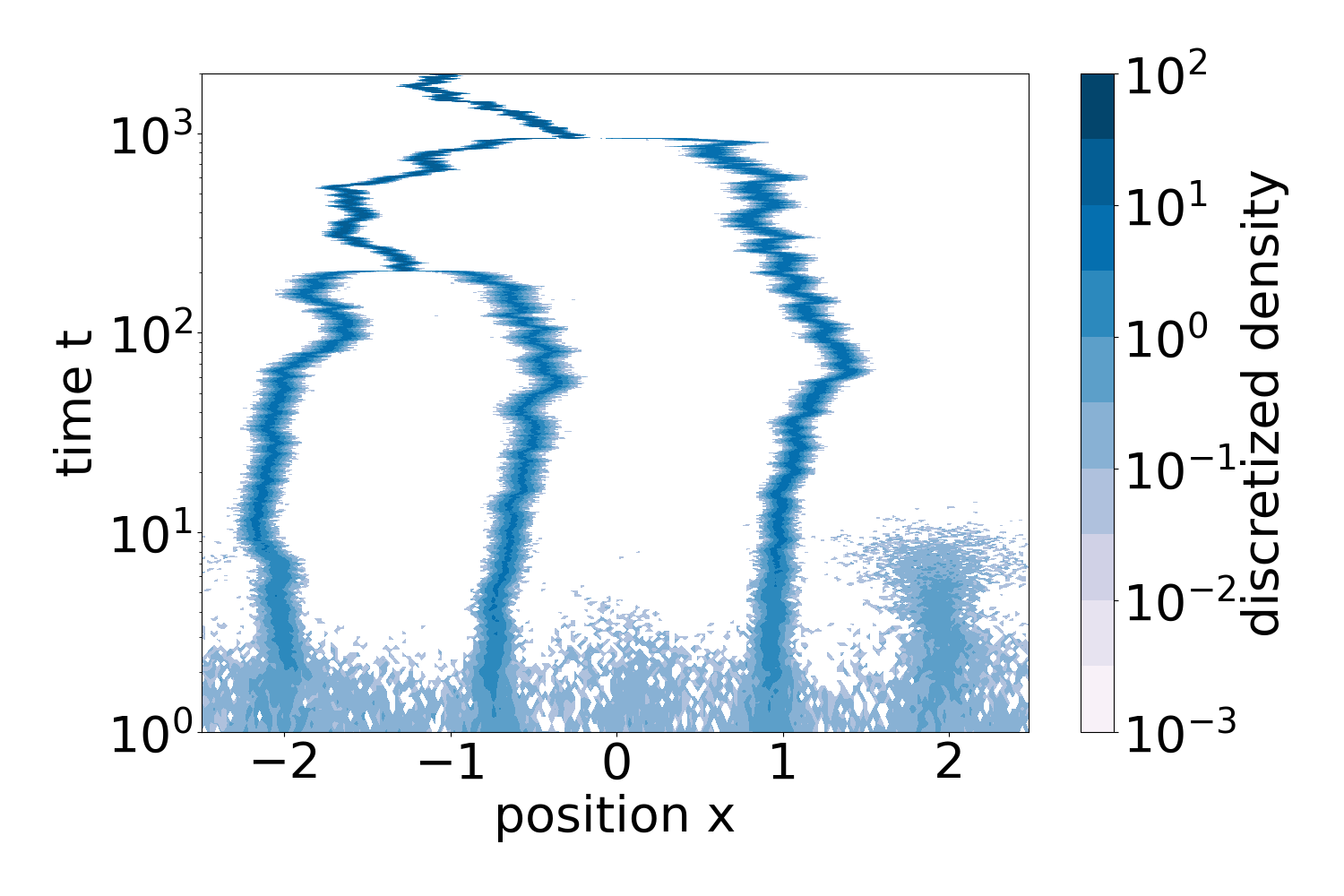}
        \caption{empirical distribution of particle positions}
        \label{fig:PBD_compared_SPDE_long_PBD}
    \end{subfigure}
        \begin{subfigure}{0.49\textwidth}
    \includegraphics[width=\textwidth]{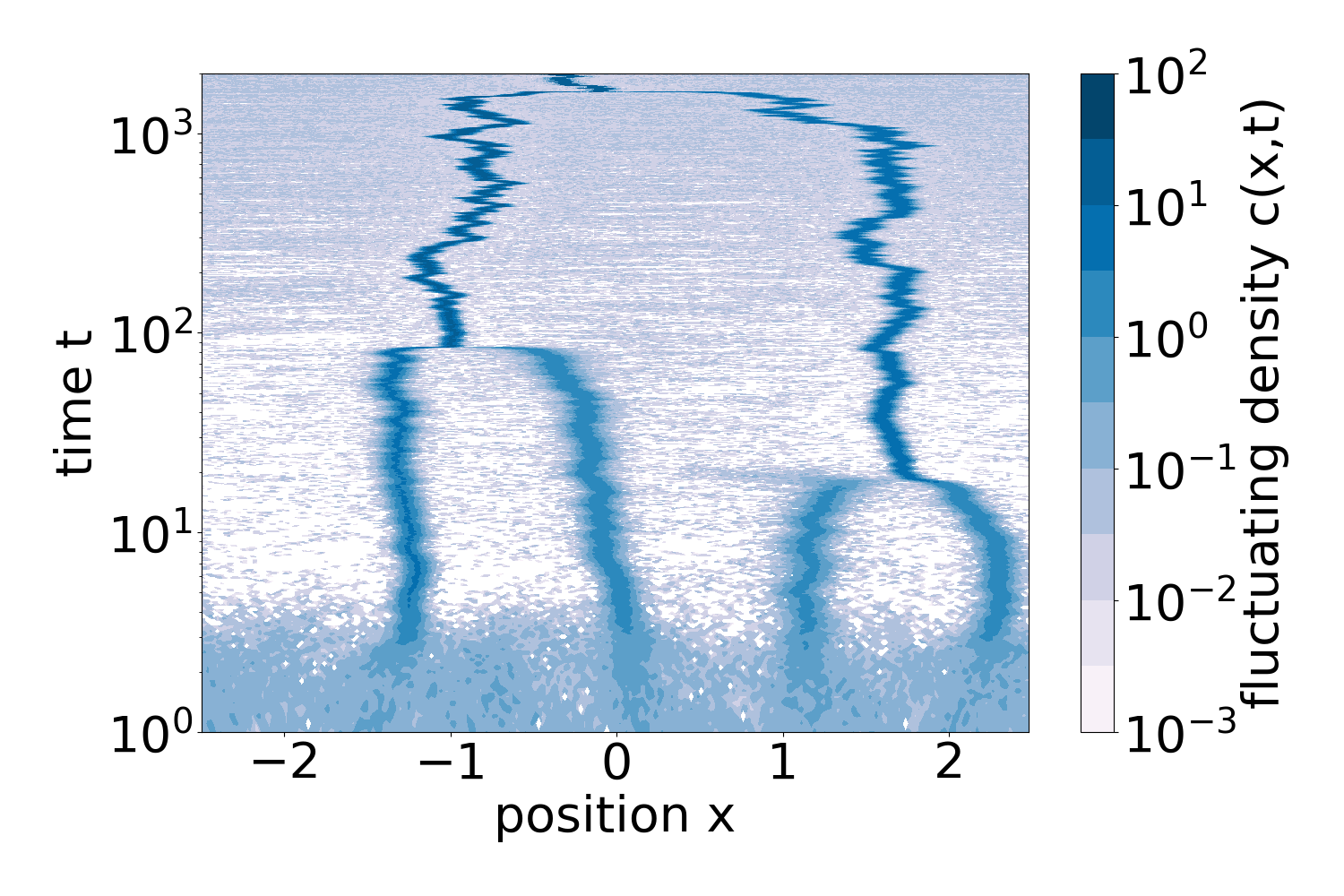}
        \caption{numerical SPDE solution}
        \label{fig:PBD_compared_SPDE_long_SPDE}
    \end{subfigure}
    \caption{\textbf{Individual long-term simulations.} Stochastic dynamics of $N=10^3$ particles  until time $t=2 \cdot 10^3$, after starting at time $t=0$ with a uniform distribution. (a)~Empirical distribution for particle-based dynamics~\eqref{eq:PBD} and (b)~simulation of the Dean--Kawasaki equation \eqref{eq:Dean-Kawasaki}. These are the same realizations as in Figure~\ref{fig:PBD_compared_SPDE}, but for a longer simulation period and on a logarithmic timescale. Please note: The two simulations, particle-based and Dean-Kawasaki-based, are subject to different realizations of the respective noise process and thus cannot be compared directly.}
    \label{fig:PBD_compared_SPDE_long}
\end{figure}

Figure~\ref{fig:PBD_compared_SPDE_long} shows individual numerical simulations of the particle-based dynamics and the SPDE solution over an extended time period. This gives an impression of the clustering dynamics in the second stage, which follows the initial cluster formation period.
For both models, the particles have aggregated into four clusters after the initial cluster formation period.
Now, the cluster centers move in space, with their movement resembling that of Brownian particles, as stated
in~\cite{garnier2016}. When getting close in space, two clusters possibly merge to one. Note that, by the logarithmic scaling of the time axis, mergers may appear like discontinuous transitions (see especially the last coalescence in Figure~\ref{fig:PBD_compared_SPDE_long_PBD}, but actually, they are continuous dynamics. 
We observe this attraction and merging for both models, with the number of clusters gradually decreasing to one. This can be interpreted as consensus convergence in the context of opinion dynamics. The waiting times for subsequent mergers and the cluster widths appear similar in both models.

We note that the depicted trajectories are individual stochastic realizations, and, naturally, other realizations will look different for each model. Nevertheless, these trajectories suggest a qualitatively high-level agreement of the spatiotemporal clustering dynamics on the particle-based and the SPDE level. For a quantitative comparison, we will examine the number of clusters evolving over time.  

\subsection{Cluster detection based on relative local maxima}
\label{subsec:Cluster detection method}

Various numerical methods are available to determine the number and position of clusters given the particles' spatial arrangement in terms of individual positions or fluctuating density. For the results presented in the main text, we used a method detecting relative local maxima in the given densities. As for the SPDE solution, clusters as local maxima can be directly identified in the fluctuating density if fluctuations of too small width are not taken into account. For the particle-based model, the relative local maxima can be found in the corresponding discretized density, which of course varies depending on the bin size (denoted by $\tilde{dx}$ in the following).

For the purpose of detecting relative local maxima, the \textit{find\_peaks\_cwt} function of the Python package \textit{scipy.signal}~\cite{du2006} was used, which performs a continuous wavelet transform over a highly fluctuating function.  
The most important parameter for this method is \textit{widths}, a set of values used for the transformation that should cover the expected width of peaks of interest. 
In our application of \textit{find\_peaks\_cwt}, we additionally delete the peaks that are smaller than the initial uniform distribution (i.e., smaller than $1/L$). 
For Figures~\ref{fig:number_cluster_average_models}-\ref{fig:MJP_distributions} we use a time step of $\tilde{dt}=1$ for cluster detection as well as the set of widths $\{10,11,\dots,14\}$ for \textit{find\_peaks\_cwt}. 

By testing different values for the method's parameter \textit{widths}, we assured ourselves that the number of clusters found with \textit{find\_peaks\_cwt} is stable so that slight changes in the range of values do not affect our results, see Appendix~\ref{app:Stability of cluster detection based on relative local maxima} with Figure~\ref{fig:stability_find_peaks_cwt}. This also allows us to use the same range of widths for the entire time horizon considered, even though the width of the clusters might change with their number. Nevertheless, we had to do simulations in advance to estimate the approximate widths of the clusters (which translates into \textit{find\_peaks\_cwt}'s parameter \textit{widths}), which could alternatively be derived analytically from the potential parameters~\eqref{eq:parameter_val} and the noise strength $\sigma$ by means of linear stability analysis, see~\cite{garnier2016}  for details.

We compared our results with another cluster detection method based on particle distances which is called HDBSCAN and is explained in more detail in the Appendix~\ref{app:Alternative cluster detection method}. Note that HDBSCAN compares individual particle positions and is thus directly applicable to the particle-based model, whereas the density of the SPDE solution must first be translated into $N$ positions. 
Analogous statistical evaluations with HDBSCAN can be found in Appendix~\ref{app:Alternative cluster detection method}. Both methods agree when choosing adequate parameter values. 

\subsection{Statistics of cluster counts} \label{sec:statistics}

Individual long-term simulations of the stochastic dynamics show that clusters merge in the course of time and do not split up again. That is, the number of clusters appears to be decreasing monotonically,
both for the particle-based model and for the SPDE, as in the exemplary simulation depicted in Figure~\ref{fig:PBD_compared_SPDE_long}. 
The timescales for cluster merging change, such that a smaller total number of clusters results in longer waiting times for another merge to occur. This is intuitive, as fewer clusters mean less frequent opportunities for merging.

\begin{figure}
    \centering
       \begin{subfigure}{0.49\textwidth}
    \includegraphics[width=\textwidth]{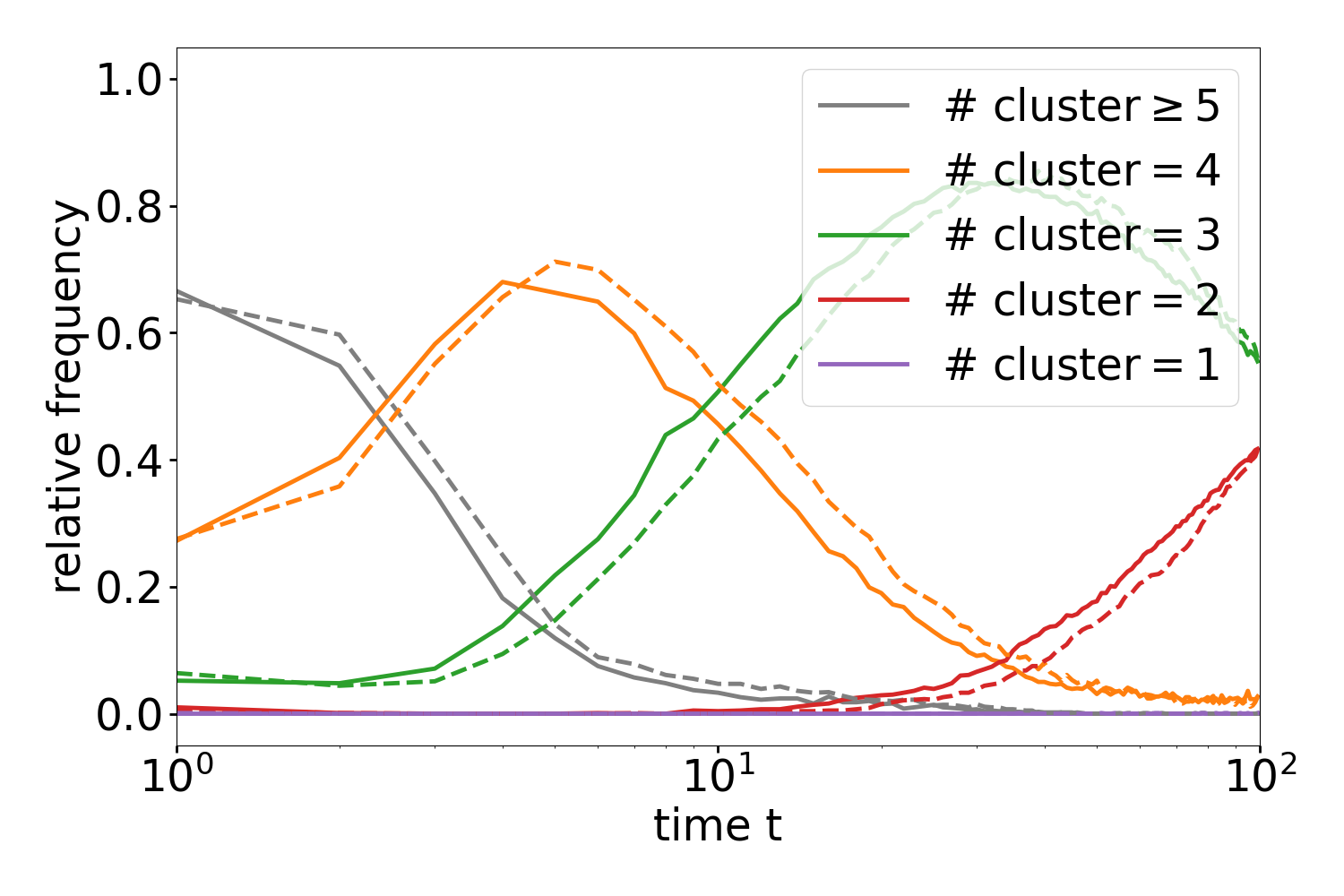}
    \caption{empirical distribution of cluster counts}
        \label{fig:number_cluster_average_models_distr}
    \end{subfigure}
     \begin{subfigure}{0.49\textwidth}
    \includegraphics[width=\textwidth]{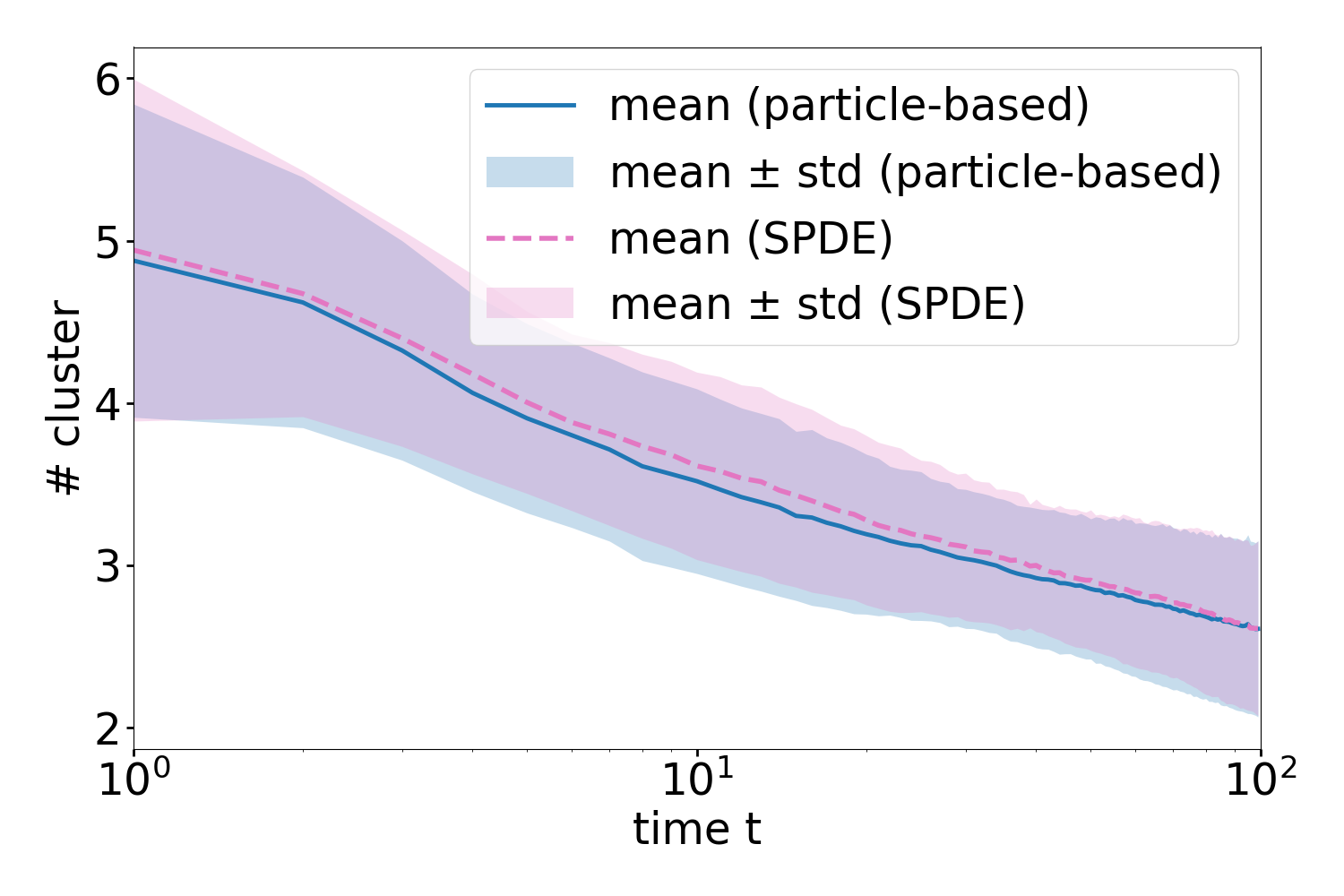}
     \caption{mean and standard deviation}
     \label{fig:number_cluster_average_models_mean}
    \end{subfigure}
    \caption{\textbf{Statistics of cluster counts over time.} (a) Relative frequencies of cluster counts and (b) associated mean and standard deviation depending on time. These results are based on $S=10^3$ independent particle-based simulations (solid lines) and $S=10^3$ independent SPDE-simulations (dashed lines), respectively, each with a population size of $N=10^3$ and starting at time $t=0$ with a uniform distribution. See Figure~\ref{fig:number_cluster_average_methods} for an analysis of the same realizations but with HDBSCAN.}
    \label{fig:number_cluster_average_models}
\end{figure}

To compare the temporal evolution of the cluster numbers for the two stochastic models, we calculate their statistical distribution, which is shown as a function of time in Figure~\ref{fig:number_cluster_average_models_distr}. The percentages of individual cluster numbers increase and decrease on logarithmic timescales, gradually merging into each other.  We observe a broad agreement between the statistics of the SPDE solution and those of the particle-based dynamics. In Figure~\ref{fig:number_cluster_average_models_mean}, the time-dependent mean and variances of the cluster counts are shown. For both models, the mean decreases monotonically over time, with the negative slope becoming flatter. Again, it is evident that the Dean--Kawasaki equation provides a very good approximation to the particle-based model.

\begin{figure}
    \centering
    \begin{subfigure}{0.32\textwidth}
    \includegraphics[width=\textwidth]{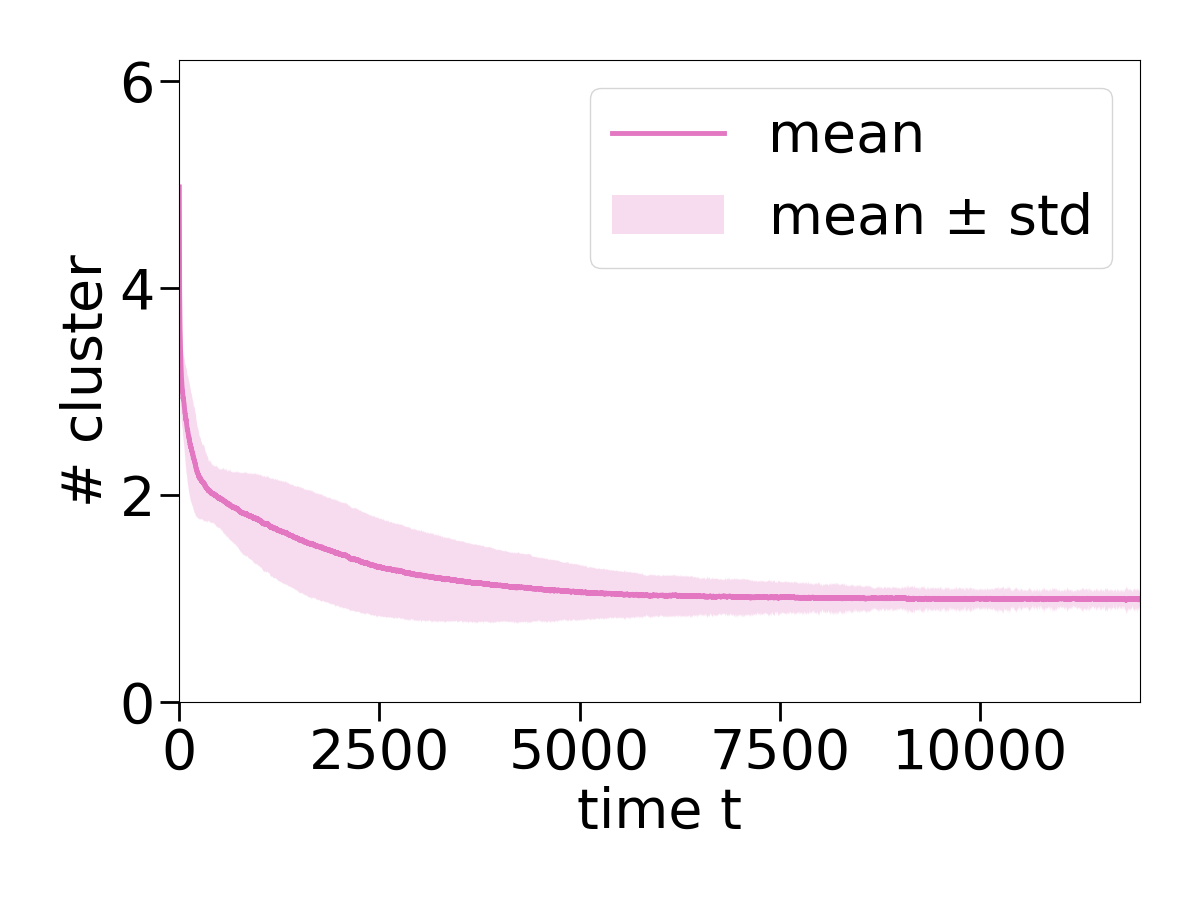}
        \caption{mean $\pm$ std}
        \label{fig:SPDE_number_cluster_average_long_mean}
    \end{subfigure}
        \begin{subfigure}{0.32\textwidth}
    \includegraphics[width=\textwidth]{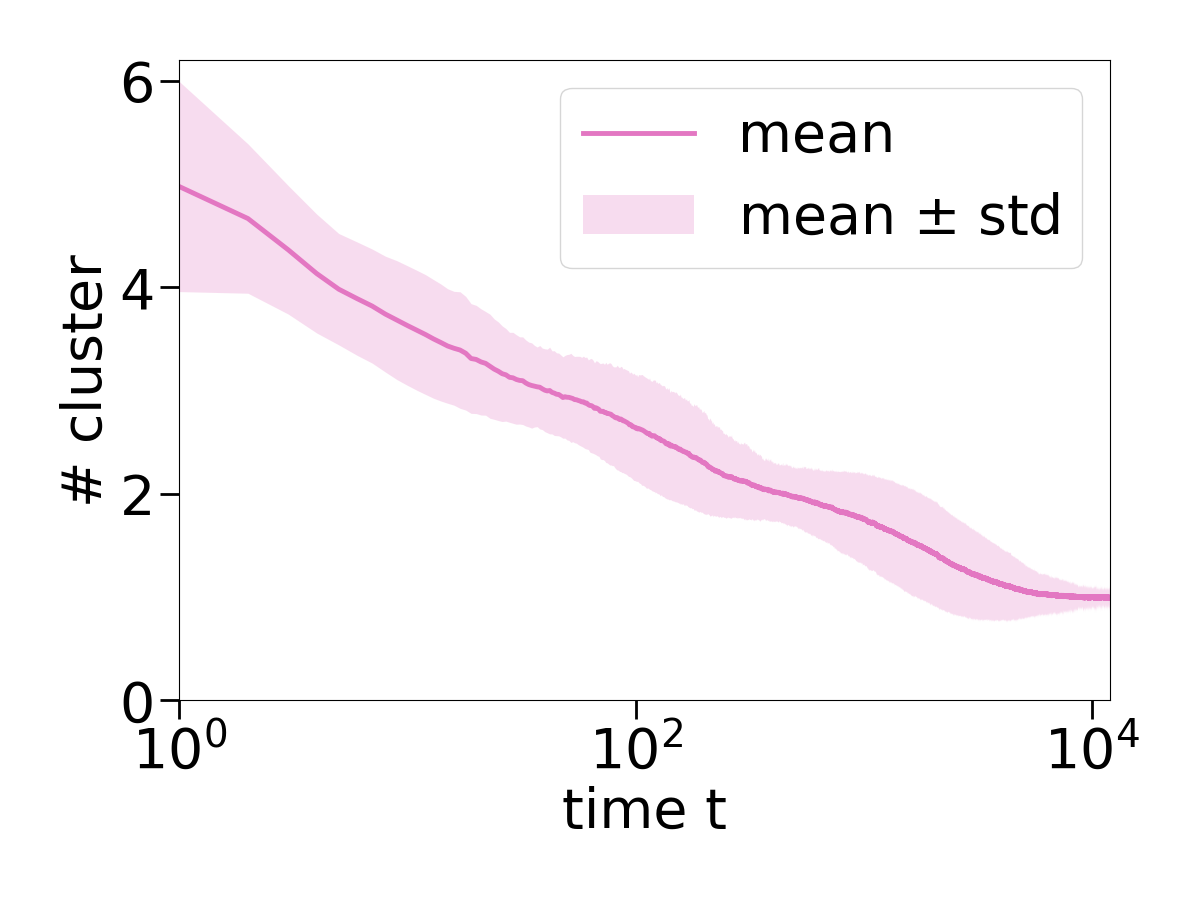}
        \caption{linear-log plot of (a)}
        \label{fig:SPDE_number_cluster_average_long_meanlog}
    \end{subfigure}
        \begin{subfigure}{0.32\textwidth}
    \includegraphics[width=\textwidth]{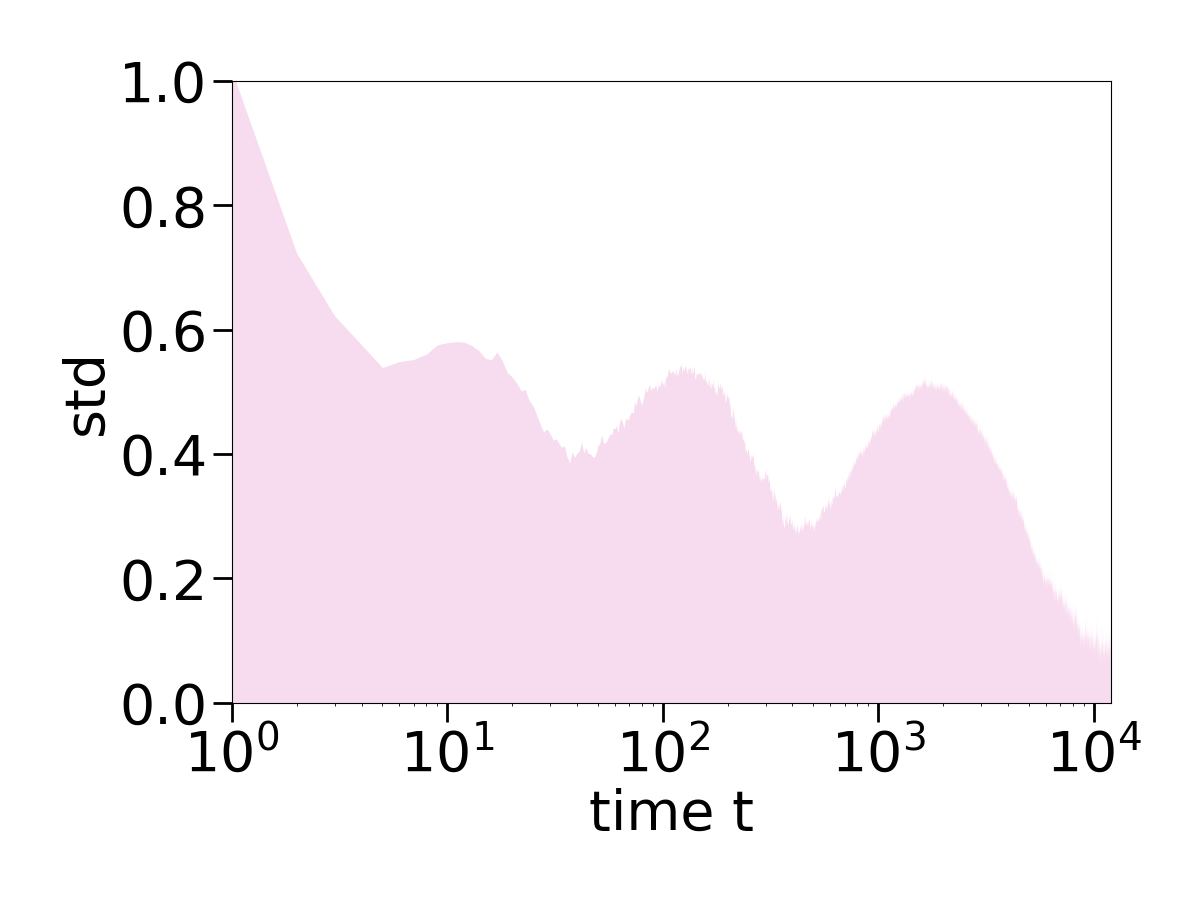}
        \caption{linear-log plot for std}
        \label{fig:SPDE_number_cluster_average_long_std}
    \end{subfigure}
    \caption{\textbf{Long-term statistics for SPDE-simulations.} Mean and standard deviation (std) of cluster counts for $S=10^3$ independent SPDE-simulations until time $t=12\cdot 10^3$ for $N=10^3$  in (a) linear scale and (b) linear-log scale after starting with a uniform distribution. (c) Standard deviation in linear-log scale. 
    }
\label{fig:SPDE_number_cluster_average_long}
\end{figure}

These insights about the conformity of the two models offer a significant advantage, as it allows for the replacement of computationally intensive particle-based simulations with faster SPDE simulations when studying the long-term evolution of the clustering process. Figure~\ref{fig:SPDE_number_cluster_average_long} shows the first- and second-order moments of the cluster counts over a fairly long period of time, estimated from SPDE-simulations. 
We observe the mean converging towards one and the standard deviation converging towards zero, see~Figure~\ref{fig:SPDE_number_cluster_average_long_mean}. The decrease in the mean number of clusters appears to be logarithmic until a certain point in time, see~Figure~\ref{fig:SPDE_number_cluster_average_long_meanlog}. 
Moreover, one can see that the standard deviation contains strong ups and downs, see~Figure~\ref{fig:SPDE_number_cluster_average_long_std}. This is due to the fact that there are time periods where many simulation runs contain cluster mergers--implying higher variability in the number of clusters--and other period where the cluster counts are more consistent between simulations.
These long-term statistics, that we can efficiently generate with the SPDE, will also be useful in the next section where we derive a Markov process for the cluster counts.

\subsection{Markov jump process for the number of clusters} \label{sec:MJP}

By comparing the first- and second-order moments of the cluster counts, as well as their time-dependent distributions, we found that the SPDE model delivers a convincing approximation of the particle-based model. We will now profit from this insight and use the computationally efficient SPDE-simulations to generate data and estimate parameter values of an even more reduced model. The idea is to define a time-homogeneous Markov jump process on $\mathbb{N}_0$ with the states referring to the number of clusters. 
This simplified model omits spatial details on cluster positions and sizes. For Markovian dynamics including these aspects, derived from mean-field PDE clustering properties, we refer to~\cite{garnier2016}. 

\subsubsection*{The cluster-counting process} More concretely, let $K(t)\in \mathbb{N}_0$ denote the number of clusters in the system at time $t$. The merging of two clusters into one corresponds to the transition $K(t) \mapsto K(t) -1$, while a split-up of one cluster into two is represented by $K(t) \mapsto K(t) +1$. The waiting times between transitions are assumed to follow time-homogeneous exponential distributions, thus neglecting the memory effects captured in spatially resolved models (particle-based and SPDE). In these models, a newly formed cluster is more likely to split again shortly after merging, as the particles have not yet fully mixed, while recently split clusters are more likely to merge again, as they have not yet moved far apart. However, as the following numerical studies will show, the Markovian \textit{cluster-counting process (CCP)} $(K(t))_{t\geq 0}$ can still provide a reasonable approximation of the cluster count distribution.

Based on linear stability analysis, see Remark~\ref{rem:LSA}, there is a time point $t_{\text{clu}}>0$ where most particles have aggregated into $n_{\text{clu}}>0$ clusters. 
We choose approximately these values as the initial time $t_0$ and state $\kappa$ for the cluster-counting process. For the parameter values under consideration and for a start in a uniform distribution, a reasonable choice is $t_0=3$, $\kappa=4$, which is in good agreement with the observations in Figure~\ref{fig:number_cluster_average_models_distr}.
Then, in our simulations, clusters merge successively, while a reverse split-up does (almost) never appear.
Based on these observations, we reduce the system state for the cluster-counting process to $\mathbb{S}:=\{\kappa,\kappa-1,...,2,1\}$ and assume the matrix $\Lambda= (\lambda_{i,j})_{i,j\in \mathbb{S}}$ of transition rates $\lambda_{i,j}\geq 0$ to have the form 
\begin{equation}
    \Lambda = \begin{pmatrix} 
    - \lambda_{\kappa,\kappa-1} & \lambda_{\kappa,\kappa-1} & 0 & \cdots & \cdots & 0 \\ 
    0 & - \lambda_{\kappa-1,\kappa-2} & \lambda_{\kappa-1,\kappa-2} & 0 & \cdots & 0 \\ \cdots & \cdots & \cdots & \cdots & \cdots & \cdots \\ 0 & \cdots & \cdots & 0 & -\lambda_{2,1} & \lambda_{2,1} \\ 0 & \cdots & \cdots & \cdots & \cdots & 0
    \end{pmatrix},
\end{equation}
where $\lambda_{i,j}\geq 0$ is the (time-independent) rate for the number of clusters $K(t)$ to switch from $K(t) =i$ to $K(t) =j$. For the initial distribution vector $\mu(t_0)=(\mu_{\kappa}(t_0),\dots,\mu_1(t_0))$ we assume that we almost surely start with $\kappa$ clusters, i.e., in $(1,0,\dots, 0)$, at time $t_0$. 
These definitions can easily be extended to another initial distribution and more general dynamics (including, e.g., split-ups of clusters). 
Given the initial distribution $\mu(t_0)$ and the transition rate matrix $\Lambda$, the distribution of the cluster-counting process at time $t> t_0$ is given by 
\begin{equation} \label{eq:mu}
    \mu(t) = \mu(t_0) e^{\Lambda(t-t_0)},
\end{equation}
with $\mu_i(t)$ being the probability to have $i$ clusters at time $t$.

\begin{figure}
    \centering
    \begin{subfigure}{0.99\textwidth}
    \includegraphics[width=\textwidth]{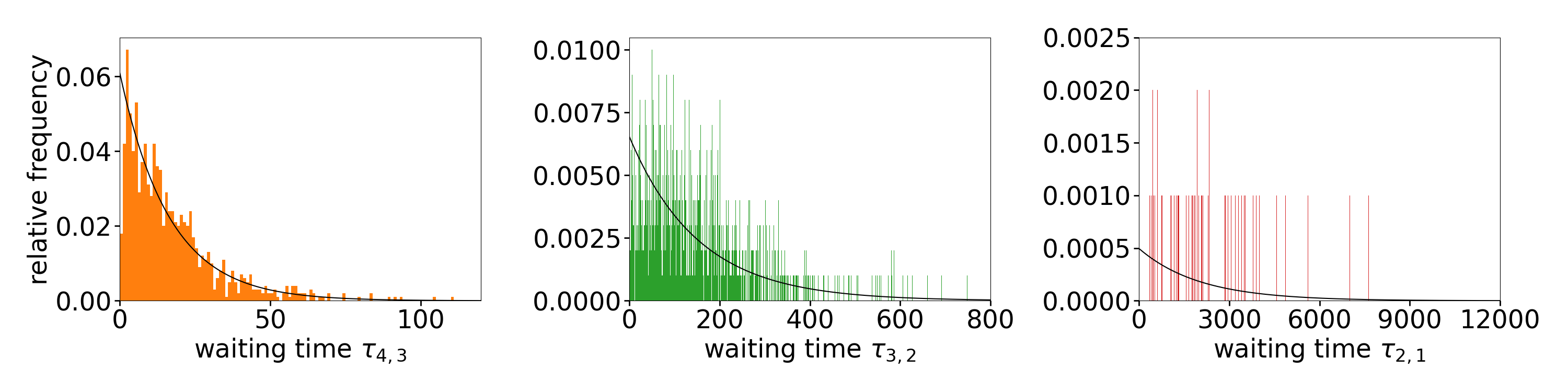}
    \end{subfigure}
    \caption{\textbf{Distribution of waiting times for cluster merge.} Relative frequencies of waiting times $\tau_{i,i-1}$ to jump from $i$ to $i-1$ clusters, see~\eqref{tau}, computed from $S=10^3$ independent SPDE-simulations with $N=10^3$ until time $t=12 \cdot 10^3$ after starting in $t=0$ with a uniform distribution, compared to best approximation by exponential distributions (black lines) with mean $\Bar{\tau}_{i,i-1}$.
    }
\label{fig:SPDE_number_cluster_waiting_times}
\end{figure}

\subsubsection*{Estimation of the jump rates} 
Let 
\begin{equation}\label{tau}
    \tau_{i,i-1}\geq 0, \quad i \in \{\kappa,\kappa-1,...,2,1\},
\end{equation}
denote the random waiting time for the cluster-counting process $K(t)$ to switch from state $i$ to state $i-1$. 
We have the relation $\mathbb{E}(\tau_{i,i-1})=1/\lambda_{i,i-1}$, which motivates to estimate the jump rates via
\begin{equation}\label{eq:bar_lambda}
    \bar{\lambda}_{i,i-1} =  (\bar{\tau}_{i,i-1})^{-1},
\end{equation}
where $\bar{\tau}_{i,i-1}\approx \mathbb{E}(\tau_{i,i-1})$ is the mean waiting time over all Monte Carlo simulations of the spatially resolved models (particle-based or SPDE).

Within the limited simulation time $T$ not all realizations necessarily reach the state of $K(t)=1$ clusters. That is, statistically we can only estimate
\begin{equation}\label{eq:cond_exp}
    \mathbb{E}(\tau_{i,i-1}\, |\,  Z_{i, i-1} \leq T),
\end{equation}
where $Z_{i,i-1} >0$ denotes the random variable modeling the time when the transition from $i$ to $i-1$ clusters occurs. For $T\to \infty$, we have $\mathbb{E}(\tau_{i,i-1}\, |\,  Z_{i, i-1} \leq T) \to \mathbb{E}(\tau_{i,i-1})$, that is, we can avoid calculating conditional expectations~\eqref{eq:cond_exp} by simulating over a sufficiently long time period. For the Dean--Kawasaki equation, this is achievable, see Figure~\ref{fig:SPDE_number_cluster_waiting_times} for the results. For particle-based simulations, however, the estimation of the rates using long-term simulations can indeed be computationally unfeasible. This highlights the benefit of employing the SPDE approach, as it allows us to gather enough data for a reliable parameter estimation. 

\begin{figure}
    \centering
    \begin{subfigure}{0.8\textwidth}
    \includegraphics[width=\textwidth]{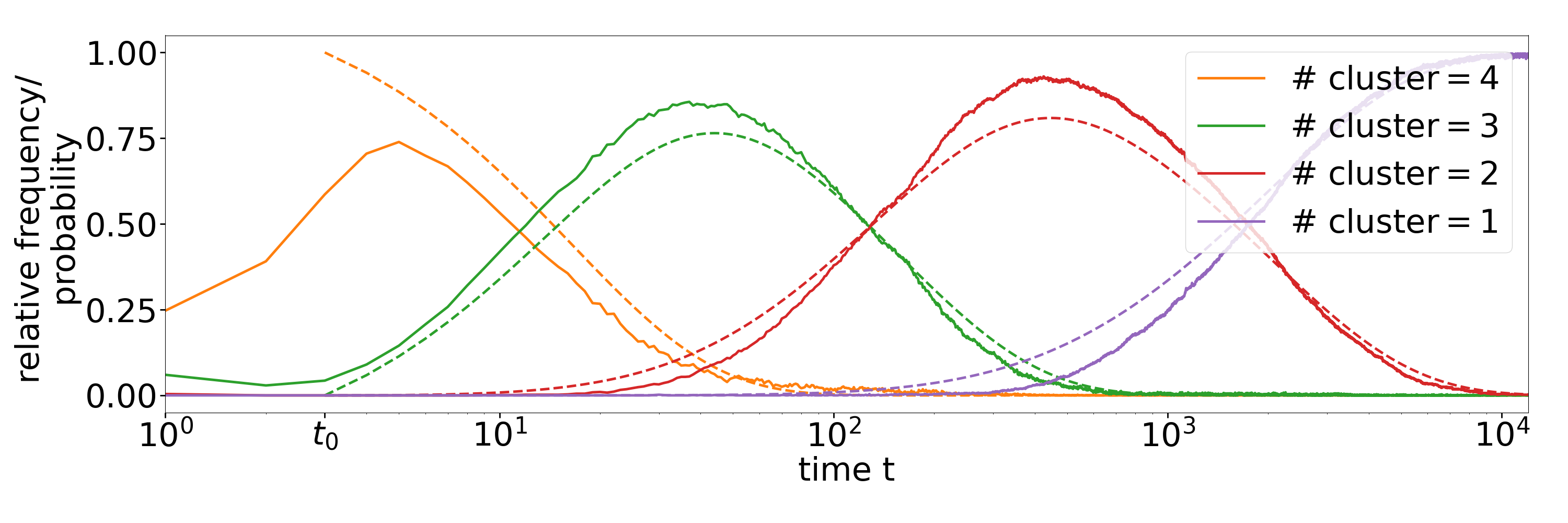}
    \end{subfigure}
    \caption{\textbf{Cluster count distribution over time: SPDE vs. CCP.} Time-dependent statistical distribution (solid) of the number of clusters computed from $S=10^3$ independent SPDE-simulations with $N=10^3$ until time $t=12 \cdot 10^3$ after starting in $t=0$ with a uniform distribution, compared to distribution $\mu(t)$ (dashed) of cluster-counting process $K(t)$, analytically determined using~\eqref{eq:mu}. For the initial distribution $\mu(t_0)$ of the CCP in $t_0=3$, we assume that the process almost surely starts with $\kappa=4$ clusters. 
    The CCP rates $\lambda_{i,i-1}$ were calculated from the statistical mean waiting times, see~\eqref{eq:bar_lambda} and Figure~\ref{fig:SPDE_number_cluster_waiting_times}. 
    }
    \label{fig:MJP_distributions}
\end{figure}

\subsubsection*{Evaluation} Given the estimated jump rates, we can analytically derive the time-dependent distribution of the cluster-counting process. This distribution is compared in Figure~\ref{fig:MJP_distributions} with the statistical distribution of cluster numbers obtained from SPDE simulations. The results show a strong qualitative agreement between the models, particularly in terms of the temporal evolution of the cluster count. 
Both models exhibit temporal peaks in the distributions, where one cluster count predominates over the others. These peaks replace each other over time. Periods of 'overlap'--where multiple cluster counts have nonzero probabilities--alternate with periods of 'dominance', where a single cluster count prevails. The timing of these periods aligns well between the two models, and the peak heights are comparable, although the SPDE values tend to be slightly higher than those of the cluster-counting process. Given that we trust the SPDE simulations to reliably capture the particle-based dynamics, our observations also suggest a strong agreement between the cluster-counting process and the particle-based dynamics.

The matching between the cluster counting process and the SPDE (resp. particle-based model) may seem surprising, given that the Markovian dynamics ignore the memory effects inherent in the spatially resolved models. This agreement can be (at least partially) attributed to the fact that cluster splitting is extremely rare in the scenario under consideration. As a result, the effects of re-splitting after merging, or re-merging after splitting, become negligible. 
Moreover, the waiting times between events scale very differently ($\tau_{4,3} \ll \tau_{3,2} \ll \tau_{2,1}$, see Figure 5), making the mean waiting time the dominant factor, while the exact distribution of the waiting times is of minor importance.

\section{Discussion and conclusion}

In this work, we showed that model approximations in the form of SPDEs are able to adequately reproduce spatiotemporal clustering dynamics observed for stochastic particle-based dynamics. 
We presented a simplified particle-based model for the diffusion and pairwise interaction of particles on a membrane, the latter represented by a one-dimensional domain. 
In the model, particles randomly move in space and, once they find each other, tend to stay close to each other due to attracting forces. Despite its simplicity, this model exhibits intriguing clustering effects, both for the initial cluster formation period and for the long-term clustering dynamics, which involves the movement and merging of clusters. While the initial period has been studied in the literature using deterministic mean-field approximations, this work focused on stochastic approaches for modeling the subsequent clustering dynamics.
Our numerical simulations showed a strong qualitative agreement between the SPDE dynamics and the particle-based dynamics regarding the number, the width, and the spatial movement of clusters.
Both models revealed a gradual merging of clusters caused by random motion in space and attraction. 
The number of clusters progressively decreased and the waiting times for each additional merger increased.
A quantitative comparison of the two models was done in terms of cluster counts statistics. We estimated the time-dependent distribution of the cluster counts, as well as their mean and variance, again finding convincing agreement of the SPDE dynamics with the underlying particle-based model. We conclude that SPDEs can significantly better reproduce long-term cluster dynamics compared to PDE approximations, which ignore stochastic effects that can be essential for clustering.

The computational efficiency of the SPDE approach allowed us to perform long-term simulations of the spatiotemporal clustering process for studying its statistics. We used the simulation data in combination with insights from linear stability analysis to estimate the parameter values and the starting point of a further reduced model, given by a Markov jump process for the cluster counts. This process ignores spatial information and memory effects, assuming exponentially distributed waiting times between two cluster mergers. Despite this radical simplification, the cluster-number distribution of the Markov process matched well with that of the SPDE process. This parameter estimation, which is based on extensive simulation data, is an example of how one can benefit from the efficient SPDE approach. 
Similarly, the SPDE can be employed to investigate other quantities of interest, such as the temporal evolution of cluster positions and widths, or boundary effects of the domain, as well as for parameter studies.

We selected the Morse potential for the interaction forces because it closely resembles membrane-mediated interactions observed in previous studies. 
However, our results will also apply to other types of attractive interactions. 
Future research could explore how our findings depend on the chosen parameter values or on model variations and extensions, including, e.g., confining potentials for the spatial movement of particles on a non-periodic domain. Another obvious next step is to extend our analysis to two-dimensional domains, thereby getting closer to realistic scenarios of receptor dynamics on membranes. For simulating the related Dean-Kawasaki equation one could use the multilevel Monte Carlo methods proposed in~\cite{cornalba2024multilevel}.  We will compare our results with the physics-based model of protein aggregation~\cite{sadeghi2021, sadeghi2022} and with experimental data~\cite{moeller2020}. Parameter estimations will be necessary to adapt our simplified model to the existing results and shed more light on the phenomenon of receptor clustering. To do this, we need to infer the potential parameters and the noise strength from data, for which we can use, for instance, the \textit{method of moments} as in~\cite{pavliotis2024, comte2024nonparametric}.

Another area in which the presented model has the potential to outshine other techniques is the possibility of including multi-body effects. Membrane-mediated interactions can generally have a multibody nature and cannot be expressed as sums of pairwise interactions between particles~\cite{fournier2024}. In the SPDE representation, by making the potential coefficients $C_r$ and $C_a$ explicit functions of particle densities, such effects can be modeled and the resulting nontrivial clustering dynamics can be studied. Furthermore, our model can serve as a basis for constructing efficient hybrid models that combine particle-based dynamics with (S)PDEs as, e.g., in~\cite{montefusco2024partial,donev2009}. 

\section*{Code availability}
The Python code used to produce simulations and plots in this paper is available on \url{https://doi.org/10.5281/zenodo.13348715}. 

\section*{Acknowledgment}
This research has been partially funded by Deutsche Forschungsgemeinschaft~(DFG) through grant~CRC~1114 (Project~No.~235221301) and under Germany’s Excellence Strategy MATH+: Berlin~Mathematics Research Center (EXC~2046/1, Project~No.~390685689). 
GP is partially supported by an ERC-EPSRC Frontier Research Guarantee through Grant No.~EP/X038645, ERC Advanced Grant No.~247031 and a Leverhulme Trust Senior Research Fellowships, SRF\textbackslash R1\textbackslash 241055.

\bibliography{references.bib}
\bibliographystyle{abbrvnat}

\appendix
\section{Appendix}
\label{sec:Appendix}

\subsection{Numerical discretization of the SPDE}\label{app:Num_SPDE}
The numerical simulation of the Dean--Kawasaki equation~\eqref{eq:Dean-Kawasaki} is performed using the finite difference scheme described in this section. For the deterministic mean-field part of \eqref{eq:Dean-Kawasaki} we use the explicit midpoint method, while treating the noise term as an explicit term like in the first time step of the simulation procedure in~\cite{cornalba2023}. 
Concretely, the torus $\mathbb{T}\subset \mathbb{R}^d$ (as a generalization of the one-dimensional domain defined in the main document) is divided into a finite set $G_h\subset \mathbb{R}^d$ of grid points using a regular mesh size $h>0$. We write $c_h^n=(c_h^n(y))_{y\in G_h}$ for the discrete approximation of the function $c$, that is, $c_h^n(y) \approx c(y,n\cdot dt)$ for a time step $dt>0$. Letting
$dW^{n}(y)=\left(dW^{n}_1(y),...,dW^{n}_d(y)\right)$, $n\in\mathbb{N}$, denote a $d$-dimensional vector of independent normal distributions, $dW^{n}_i(y)\sim \mathcal{N}(0,dt)$ for $i=1,...,d$, $y\in G_h$, the discrete scheme is given by
\begin{align}
    c^{n+1}_h&=c^n_h+dt f\left(c^n_h+\frac{dt}{2} f(c^n_h)\right) +\frac{\sigma}{\sqrt{N}} \sum_{y\in G_{h}} \nabla_h \cdot \left(\sqrt{c^{n}_{h+}} \, e_y\right)dW^{n+1}(y) \label{eq:noise_term}
\end{align}
for $c^n_{h+}:=\max\{0,c^n_h\}$, $e_y(x):=h^{-d/2}\delta_{yx}$
and
\[f: \mathbb{R}^m \to \mathbb{R}^m,\quad  f(z):=\frac{\sigma^2}{2} \Delta_h z +\nabla_h (z \cdot(\nabla_h F_h*z)),\]
where $z=(z(y))_{y\in G_h}$.
Here, $\nabla_h=(\nabla_{h,x_1},...,\nabla_{h,x_d})$ is the associated finite difference gradient operator and $\Delta_h$ is the discrete Laplace operator. Moreover, the symbol $*$ stands for the discrete convolution operator (i.e., a sum instead of an integral), and $F_h$ is defined by $F_h:=(F(y))_{y\in G_h}$.

For $d=1$ we can write
\begin{equation} \label{eq:grid}
  G_h=\left\{-\frac{L}{2},-\frac{L}{2}+h, \dots, \frac{L}{2}-h, \frac{L}{2}\right\} 
\end{equation}
for $h=L/(m-1)$, where $m=|G_h|$ is the number of grid points. For the discrete gradient operator we choose central differences:
\[ \nabla_h: \mathbb{R}^m \to \mathbb{R}^m,\quad (\nabla_h z)(y) = \frac{z(y+h)-z(y-h)}{2h}. \]
The discrete Laplacian is given by
\[ \Delta_h: \mathbb{R}^m \to \mathbb{R}^m,   \quad  (\Delta_h z)(y)= \frac{z(y+h)-2z(y)+z(y-h)}{h^2}.\]
For the one-dimensional case, the noise term in Eq.~\eqref{eq:noise_term} reads 
\[
   \frac{\sigma}{\sqrt{N}}\frac{1}{2h\sqrt{h}} \left[\sqrt{c^n_{h+}(y+h)}dW^{n+1}(y+h)-\sqrt{c^n_{h+}(y-h)}dW^{n+1}(y-h)\right].
\]

\subsection{Stability of cluster detection based on relative local maxima}\label{app:Stability of cluster detection based on relative local maxima}

\begin{figure}
    \centering
     \begin{subfigure}{0.49\textwidth}
    \includegraphics[width=\textwidth]{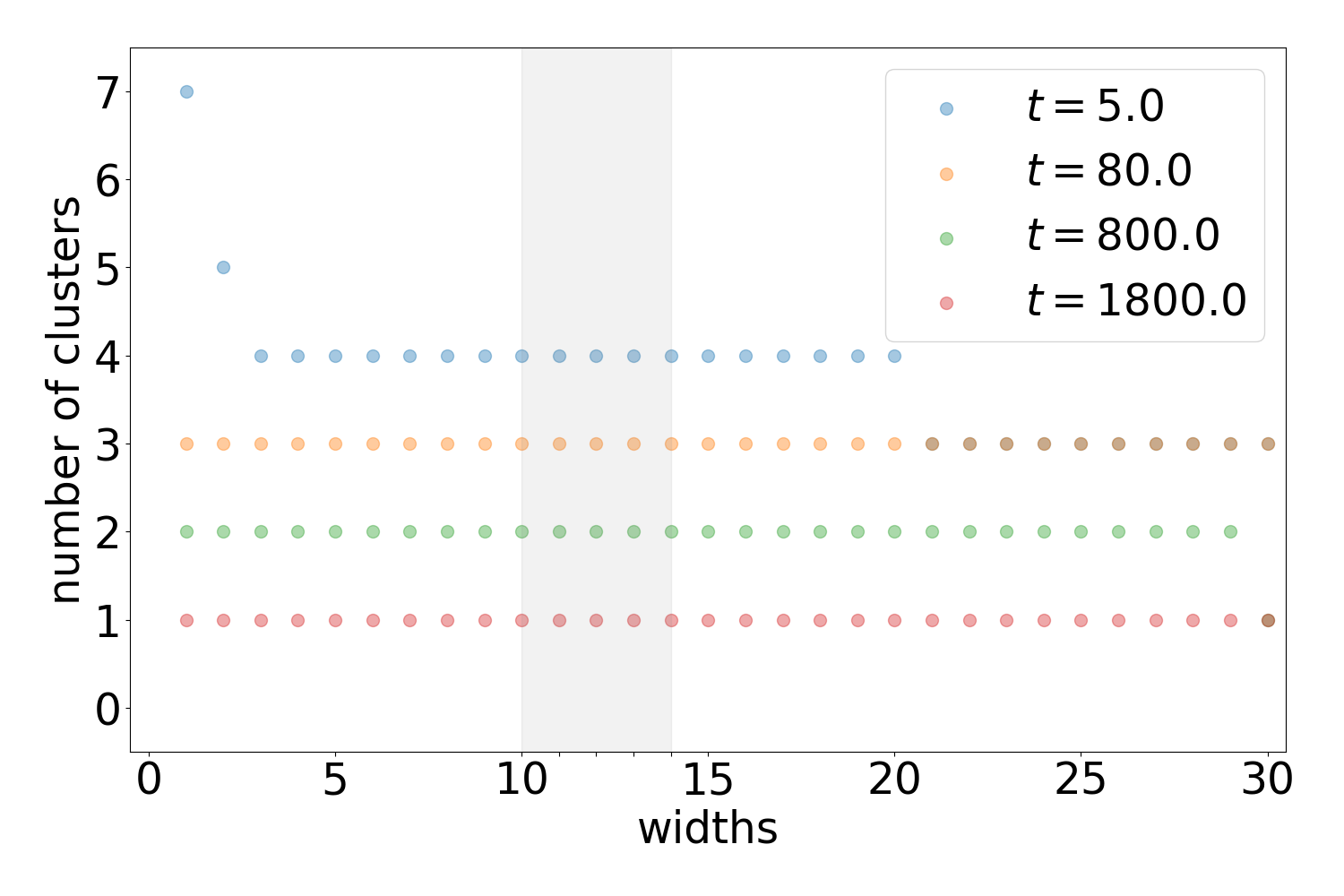}
    \caption{particle-based dynamics}
    \end{subfigure}
    \begin{subfigure}{0.49\textwidth}
    \includegraphics[width=\textwidth]{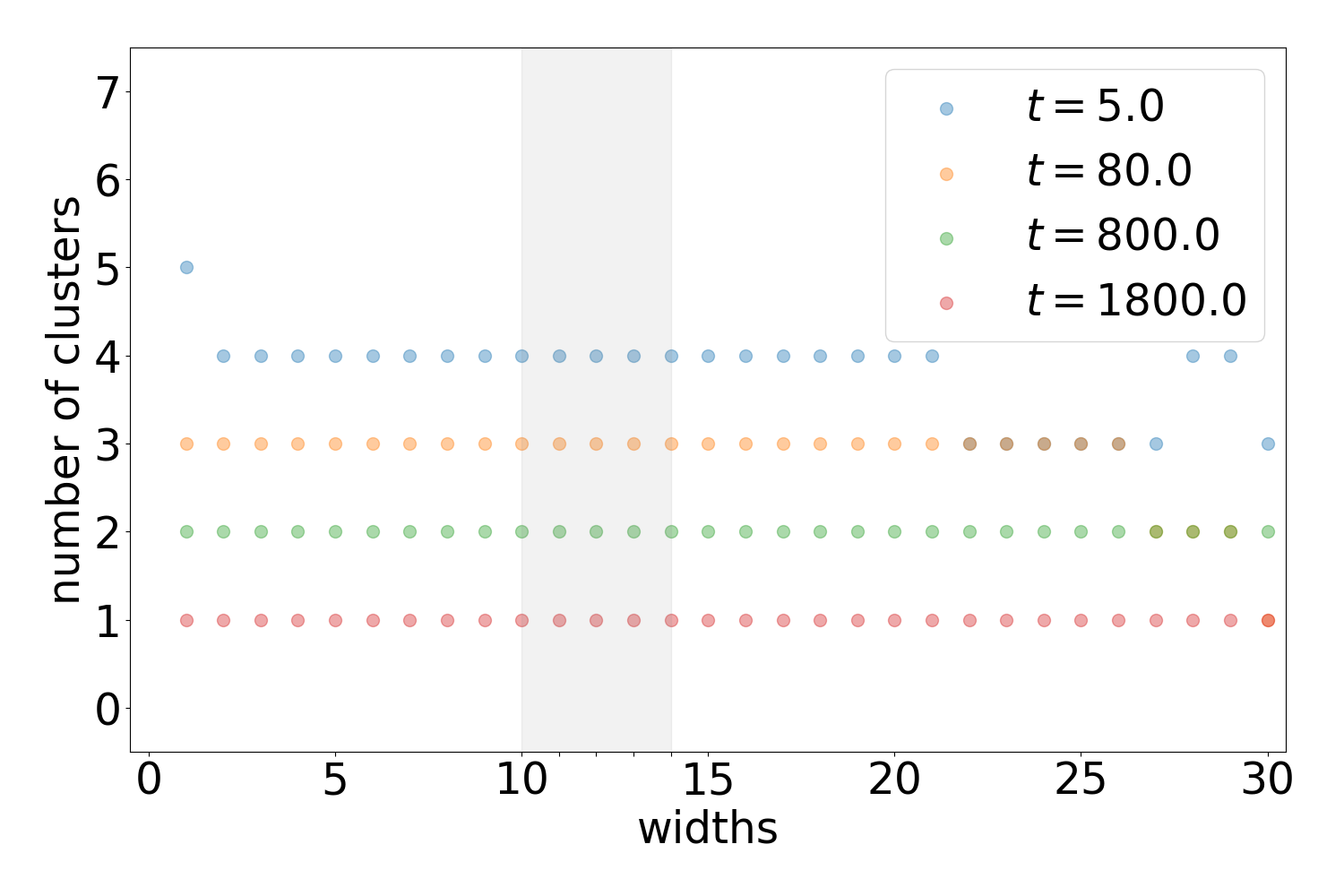}
    \caption{SPDE}
    \end{subfigure}
    \caption{\textbf{Stability of \textit{find\_peaks\_cwt}.} Stability of the number of clusters detected with \textit{find\_peaks\_cwt} against moderate changes of the value of parameter \textit{widths} (our default range is $\{10, 11, \dots, 14\}$ marked in grey). The detection method is applied to the stochastic dynamics of $N=10^3$ particles at time points $t\in\{5, 80, 800, 1800\}$, after starting at time $t=0$ with a uniform distribution. 
   (a) Particle-based dynamics~\eqref{eq:PBD}, (b) simulation of the Dean--Kawasaki equation \eqref{eq:Dean-Kawasaki}, using the same simulations as shown in Figure~\ref{fig:PBD_compared_SPDE_long}.  
The time points are chosen such that all possible numbers (and thus widths) of clusters are represented, see Figure~\ref{fig:PBD_compared_SPDE_long}.} 
    \label{fig:stability_find_peaks_cwt}
\end{figure}

In order to investigate the stability of the cluster detection method \textit{find\_peaks\_cwt} with respect to the parameter value of \textit{widths}, we plotted the number of detected clusters for different points in time (and thus different numbers and widths of clusters) in Figure~\ref{fig:stability_find_peaks_cwt}. We carry out the stability analysis for both the particle-based model and the Dean--Kawasaki SPDE. It can be seen that our chosen range of widths $\{10, 11, \dots, 14\}$ is stable for slight upward and downward changes for all chosen points in time. 
Particularly in the early stages, shortly after cluster formation, the width parameter should be carefully chosen: if set too small, minor fluctuations may be incorrectly identified as clusters, while if set too large, multiple clusters may be merged into one. With our chosen value, we position it relatively precisely in the middle of the acceptable range for $t=5$. 
The fewer clusters there are, the more stable the cluster detection becomes regarding the values of \textit{widths}.

\subsection{Cluster detection based on particle positions}\label{app:Alternative cluster detection method}

In addition to the cluster detection method using relative local maxima described in Section~\ref{subsec:Cluster detection method} (with the code \textit{find\_peaks\_cwt}), we applied the cluster detection algorithm  \textit{DBSCAN} (Density-Based Spatial Clustering of Applications with Noise)~\cite{ester1996} to ensure that our results are not method specific. In contrast to looking at particle densities to detect clusters, DBSCAN takes individual particle positions as an input, and thus is more suitable for the particle-based model. Based on the insight that points within a cluster have comparatively small distances to each other, the DBSCAN algorithm assigns individual particles to clusters by comparing the distances between their positions.
A follow-up to DBSCAN is \textit{HDBSCAN} (Hierarchical Density-Based Spatial Clustering of Applications with Noise)~\cite{campello2013}, which is more robust in terms of parameter selection. 
What is detected as a cluster by HDBSCAN depends primarily on the parameter $M^\text{min}_{\text{cluster}}\leq N$, representing the minimum number of particles required within a subset for it to be recognized as a cluster. Groups of particles smaller than $M^\text{min}_{\text{cluster}}$ are not assigned to any cluster and are returned as \textit{outliers}, which are, however, negligible for our model parameter values~\eqref{eq:parameter_val} (attractive regime, small noise), since after a short initial time almost all particles belong to a cluster. 
For applying (H)DBSCAN to the SPDE solution, the fluctuating density must first be translated into $N$ particle positions, e.g., using inverse transform sampling.

\begin{figure}
    \centering
     \begin{subfigure}{0.49\textwidth}
    \includegraphics[width=\textwidth]{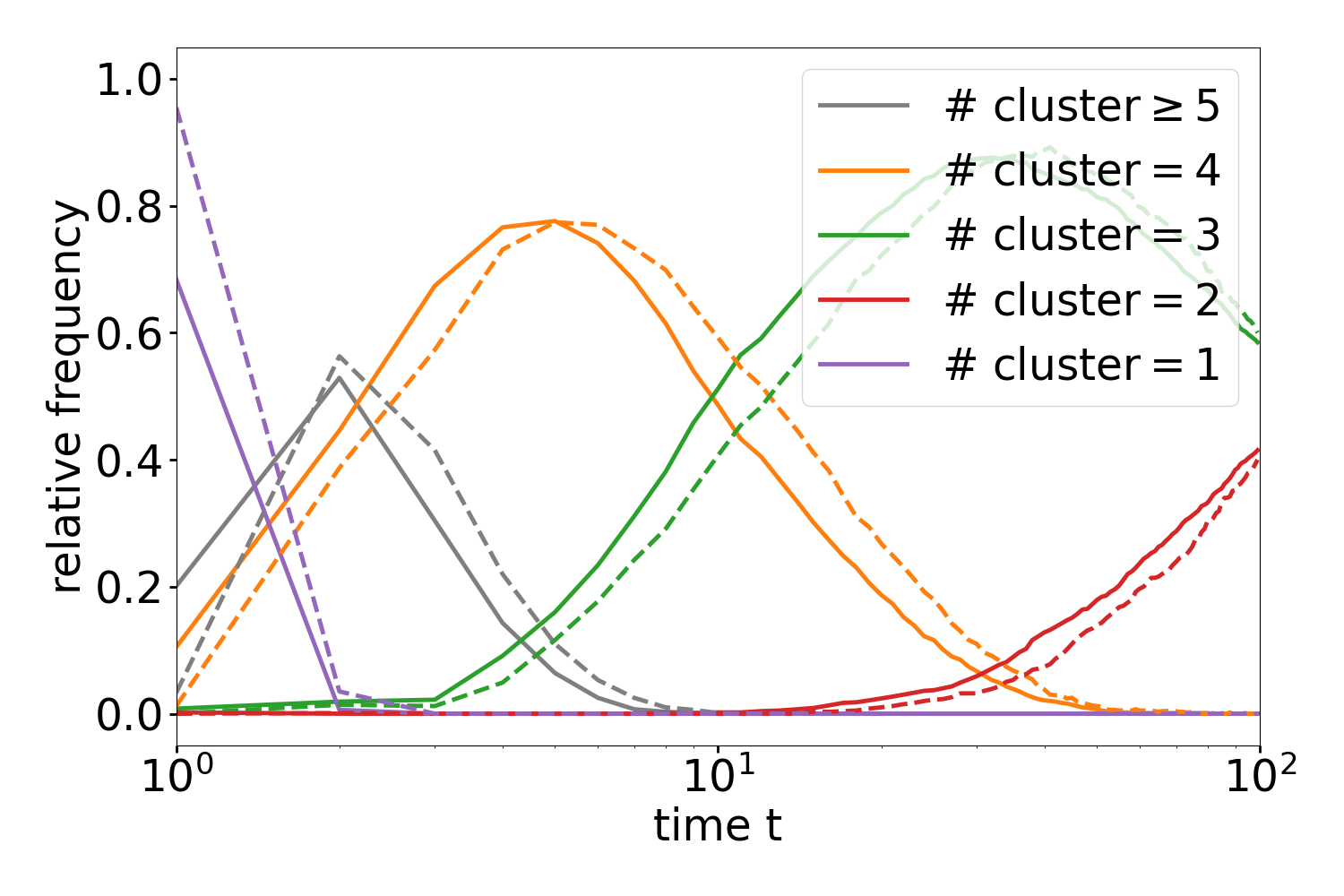}
        \caption{empirical distribution of particle positions}
        \label{fig:number_cluster_average_methods_distr}
    \end{subfigure}
    \begin{subfigure}{0.49\textwidth}
    \includegraphics[width=\textwidth]{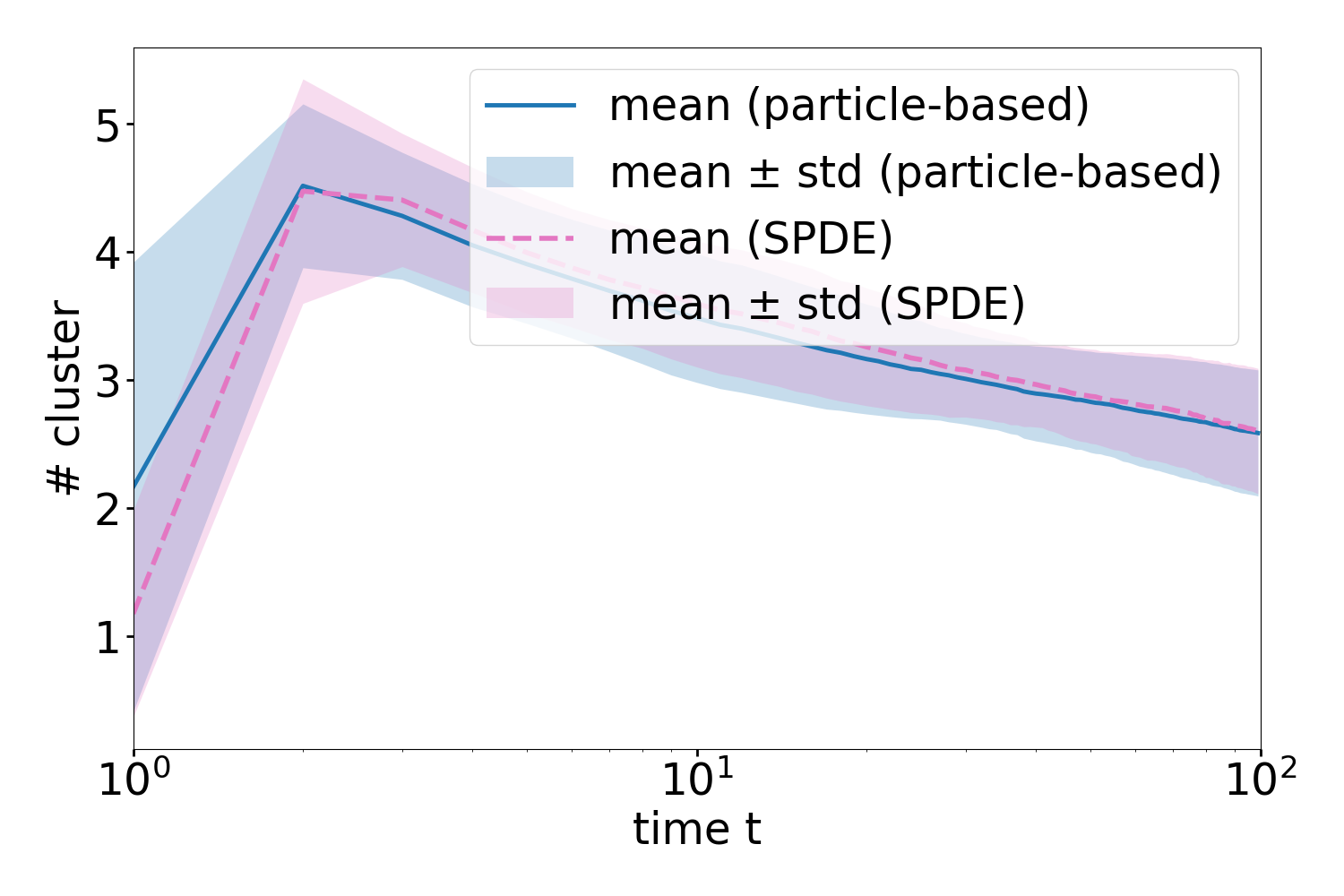}
    \caption{mean and standard deviation}
    \label{fig:number_cluster_average_methods_mean}
    \end{subfigure}
    \caption{\textbf{Statistics of cluster counts over time (HDBSCAN).} (a) Relative frequencies of cluster counts and (b) associated mean and standard deviation depending on time. These results are based on $S=10^3$ independent particle-based simulations (solid lines) and $S=10^3$ independent SPDE-simulations (dashed lines), respectively, each with a population size of $N=10^3$ and starting at time $t=0$ with a uniform distribution. Cluster detection with HDBSCAN. See Figure~\ref{fig:number_cluster_average_models} for cluster analysis of the same realizations but with \textit{find\_peaks\_cwt}.}
    \label{fig:number_cluster_average_methods}
\end{figure}

Figure~\ref{fig:number_cluster_average_methods} shows the statistical evaluations from Figure~\ref{fig:number_cluster_average_models} with HDBSCAN used instead of \textit{find\_peaks\_cwt}, confirming that both methods deliver reliable results regarding the cluster count. Here, the time step for cluster detection is $\tilde{dt}=1$ and the minimum number of particles for a cluster is $M^\text{min}_{\text{cluster}}=50$ of in total $N=10^3$ particles. 
Again, we use a time step of $dt=0.001$ for the simulations of the particle-based model as well as for the SPDE-simulations, and a spatial grid size of $h=L \cdot 2^{-8}$ for the SPDE-simulations. 

\begin{figure}
    \centering
    \begin{subfigure}{0.49\textwidth}
    \includegraphics[width=\textwidth]{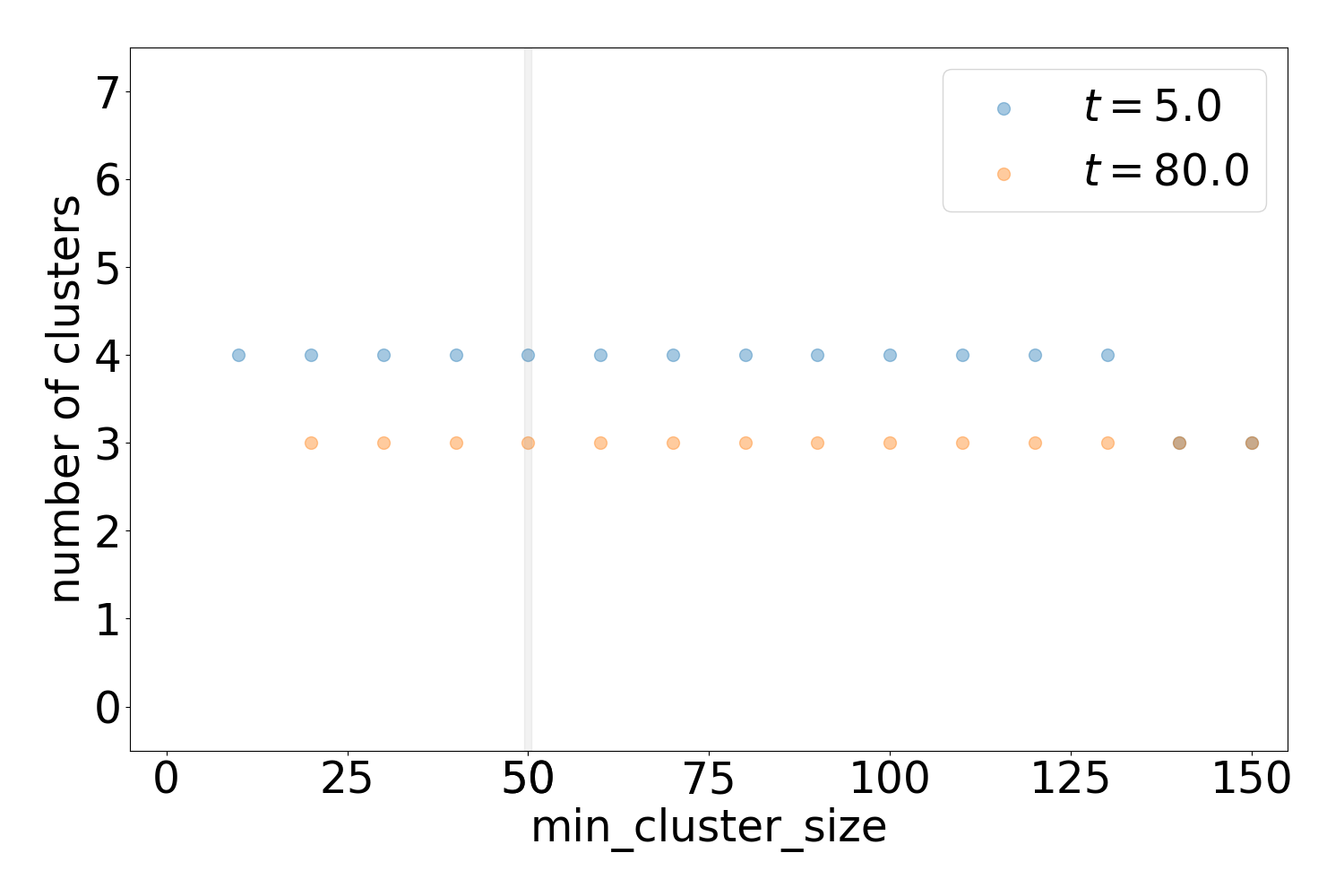}
    \caption{particle-based dynamics}
    \end{subfigure}
    \begin{subfigure}{0.49\textwidth}
    \includegraphics[width=\textwidth]{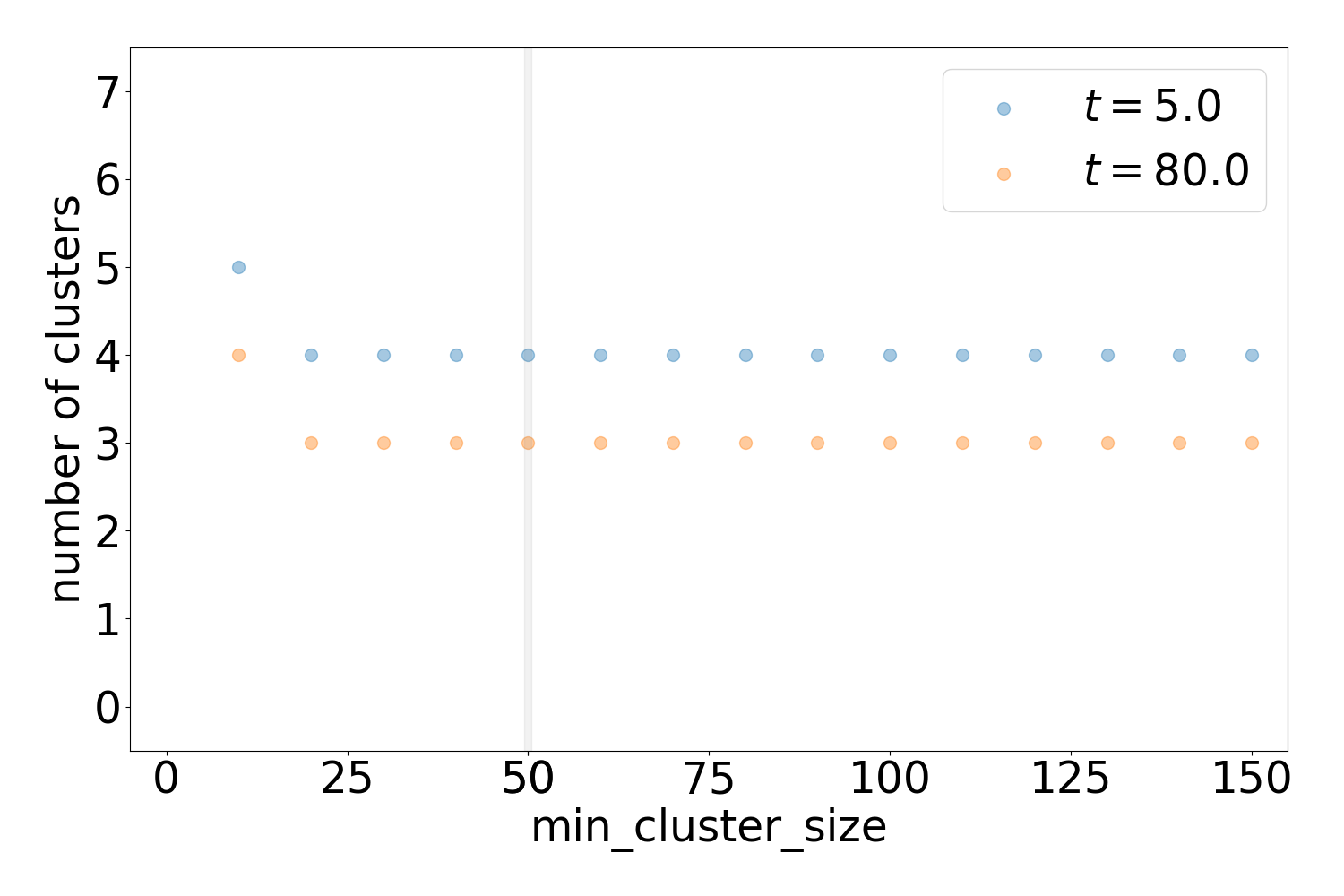}
    \caption{SPDE}
    \end{subfigure}
    \caption{\textbf{Stability of HDBSCAN.} Stability of the number of clusters detected with HDBSCAN against moderate changes of the value of parameter $M^\text{min}_{\text{cluster}}$ (our default value is $M^\text{min}_{\text{cluster}}=50$ marked in grey). Detection method is applied to the stochastic dynamics of $N=10^3$ particles at time points $t\in \{5, 80\}$, after starting at time $t=0$ with a uniform distribution. (a) Particle-based dynamics~\eqref{eq:PBD}, (b) simulation of the Dean--Kawasaki equation \eqref{eq:Dean-Kawasaki}, using the same simulations as shown in Figure~\ref{fig:PBD_compared_SPDE_long}. Note that for later time points, when there are less clusters, $M^\text{min}_{\text{cluster}}=50$ is no longer necessarily a correct or stable choice.} 
    \label{fig:stability_HDBSCAN}
\end{figure}

Just as in the application of \textit{find\_peaks\_cwt}, we observe stability regarding the choice of parameter values for HDBSCAN. Slight changes in $M^\text{min}_{\text{cluster}}$ do not alter the results during the period under consideration up to $t=100$, which we have confirmed by numerical tests, see Figure~\ref{fig:stability_HDBSCAN}. For longer periods the number of particles per cluster increases (due to merging) and larger $M^\text{min}_{\text{cluster}}$ is necessary to recognize these bigger clusters.
For the time period under consideration, the value of parameter $M^\text{min}_{\text{cluster}}$ could be determined also analytically using the parameter values~\eqref{eq:parameter_val} of the interaction potential and of the noise strength $\sigma$ for the linear stability analysis presented in~\cite{garnier2016}.

Additionally, one can observe in Figure~\ref{fig:number_cluster_average_methods} that in the initial time period of cluster formation (until $t_0$), HDBSCAN finds only one large cluster, i.e., it initially detects less clusters than \textit{find\_peaks\_cwt}, see Figure~\ref{fig:number_cluster_average_models}.

\end{document}